\documentclass[11pt, a4paper]{article}

\usepackage{fullpage}
\usepackage{amsmath, amsfonts, amssymb}
\usepackage{accents}
\usepackage{color}
\usepackage{mathrsfs}
\usepackage{graphicx}
\usepackage{multirow}
\usepackage{subfigure}
\usepackage[small,bf]{caption}
\usepackage{authblk}

\usepackage[hidelinks]{hyperref}

\usepackage[numbers,sort&compress]{natbib}

\usepackage[pagewise]{lineno}

\newcommand{\e}{\mathrm{e}}
\renewcommand{\d}{\mathrm{d}}
\newcommand{\td}[2]{\frac{\d #1}{\d #2}}

\newcommand{\pd}[2]{\frac{\partial #1}{\partial #2}}
\newcommand{\pdf}[2]{\partial #1/\partial #2}
\newcommand{\pdd}[2]{\frac{\partial^2 #1}{\partial #2^2}}

\newcommand{\etal}{\emph{et al.~}}

\newenvironment{subeq}[1]{
\begin{subequations}
#1
\end{subequations}
}

\title{Asymptotic analysis of the Guyer-Krumhansl-Stefan model for nanoscale solidification\footnote{Published in \emph{Applied Mathematical Modelling}, \url{https://doi.org/10.1016/j.apm.2018.03.026}}}

\author[1]{Matthew G.~Hennessy}
\author[1,2]{Marc Calvo Schwarzw\"alder}
\author[1,2]{Timothy G.~Myers\thanks{tmyers@crm.cat}}
\affil[1]{Centre de Recerca Matem\`atica, Campus de Bellaterra, Edifici C, 08193 Bellaterra, Spain}
\affil[2]{Departament de Matem\`atiques, Universitat Polit\`ecnica de Catalunya,  08028 Barcelona, Spain}

\begin{document}

\maketitle

\begin{abstract}

  Nanoscale solidification is becoming increasingly relevant in applications involving ultra-fast freezing processes and nanotechnology.  However, thermal transport on the nanoscale is driven by infrequent collisions between thermal energy carriers known as phonons and is not well described by Fourier's law.  In this paper, the role of non-Fourier heat conduction in nanoscale solidification is studied by coupling the Stefan condition to the Guyer--Krumhansl (GK) equation, which is an extension of Fourier's law, valid on the nanoscale, that includes memory and non-local effects.  A systematic asymptotic analysis reveals that the solidification process can be decomposed into multiple time regimes, each characterised by a non-classical mode of thermal transport and unique solidification kinetics.  For sufficiently large times, Fourier's law is recovered. The model is able to capture the change in the effective thermal conductivity of the solid during its growth, consistent with experimental observations.  The results from this study provide key quantitative insights that can be used to control nanoscale solidification processes.

\end{abstract}

\section{Introduction}

Advances in the field of nanotechnology are improving the efficiency, functionality, and
cost-effectiveness of modern devices. Nanowire-based solar cells, for instance, offer several
advantages over traditional wafer-based and thin-film technologies \cite{garnett2011}. Furthermore,
the unique physical properties of carbon nanotubes have enabled the fabrication of
new electrochemical biosensors \cite{wang2005}.
Nanotechnology is also playing an increasing role in biology and medicine \cite{salata2004}, 
where it finds applications in drug and gene delivery \cite{mah2000}, 
protein detection \cite{nam2003}, and tissue engineering \cite{ma2003}.
A key issue surrounding the use of nanoelectronic devices \cite{pop2006},
nano-enabled energy systems \cite{poudel2008}, and nanomedicine \cite{hamad2002} is that of
thermal management \cite{cahill2003, siemens2010}. The ability to successfully manipulate heat can 
be vital to device performance \cite{cahill2014} and a lack of thermal regulation can lead to melting
and device failure \cite{nie2011}. Understanding nanoscale heat transfer and phase change is
therefore crucial for current and future applications of nanotechnology.

At the nanoscale, heat transfer and phase change become markedly different from their macroscopic
counterparts. This is partially attributed to the increased ratio of surface-to-bulk atoms,
which can introduce a size dependence to key thermodynamic parameters such as melt temperature
\cite{buffat1976,david1995,wronski1967}, latent heat \cite{lai1996, zhang2000, sun2007}, and surface
energy \cite{tolman1949}. Furthermore, the mean free path of thermal energy carriers, known as phonons,
can be on the order of hundreds of nanometers in crystalline solids at room 
temperature \cite{siemens2010}. As a result, thermal transport on the nanoscale occurs as a 
ballistic process that is driven by infrequent collisions between
phonons, in contrast to macroscopic thermal transport, which is a diffusive process
driven by frequent collisions and gradients
in the temperature. The ballistic nature of nanoscale heat transport can lead to substantial
decreases in the effective thermal conductivity of nanomaterials, 
which has been experimentally confirmed \cite{chang2008, li2003, siemens2010} in samples with length
scales up to 10 microns \cite{johnson2013}, far beyond the nano-regime.

Extensive research has been carried out to develop practical theories of heat transport and
phase change that are valid at the nanoscale. The role of size-dependent parameters in nanoparticle 
\cite{back2014, font2013, font2015, mccue2008, myers2015, ribera2016} and nanowire \cite{florio2016, goswami2010} 
melting has been studied using Fourier-based models of heat conduction \cite{myers2016}. 
However, models that are derived from Fourier's law can only capture diffusive thermal transport and
lead to an infinite speed of heat propagation,
in clear contrast to the ballistic nature of nanoscale heat transport observed in experiments. 
Several approaches have been aimed at addressing this shortcoming \cite{jou2010}. 
% Kinetic models such as
% the Boltzmann transport equation (BTE) and equation of photon radiative transport (EPRT) 
Cattaneo \cite{cattaneo1958} proposed that a temperature gradient can only induce a thermal flux after 
a finite amount of time has passed. An expansion of the governing equations about small relaxation
times leads to the hyperbolic heat equation (HHE), or Maxwell--Cattaneo equation, which captures the
wave-like propagation of heat associated with ballistic transport. Although the HHE correctly describes
heat propagation with finite speed, the introduction of a relaxation time is somewhat ad-hoc and
masks the underlying physics of nanoscale thermal transport. 
Guyer and Krumhansl \cite{guyer1966i, guyer1966ii} later derived
from the linearised Boltzmann transport equation an extension to the HHE 
which includes non-local effects and
explicitly incorporates the phonon mean free path into the governing equations. 
The Guyer--Krumhansl (GK) equation is particularly 
appealing from a theoretical point of view because it provides a link between kinetic and continuum
models and is based on well-defined physical parameters.   
Moreover, the striking similarity between the GK and Navier--Stokes equations enables
nanoscale heat transport to be conceptualised in terms of fluid mechanics
and this analogy has been used to rationalise the reduced thermal conductivity of
nanosystems in terms of phonon hydrodynamics \cite{alvarez2009, jou2010, calvo2018}.

% An extension of the HHE, which
% includes non-local effects and explicitly incorporates the phonon mean free path into the governing
% equations, was derived from the linearised Boltzmann transport equation by 
% Guyer and Krumhansl . The Guyer--Krumhansl (GK) equation is particularly
% appealing from a theoretical point of view because it has a form that is similar to the 
% Navier--Stokes equations and thus enables nanoscale heat transport to be conceptualised in terms
% of fluid mechanics \cite{jou2010, alvarez2009}.

% Ad-hoc approach that masks the underlying physics. GK equation links kinetic parameters to
% continuum parameters

Theoretical studies of nanoscale phase change that incorporate non-Fourier heat transport are
predominantly based on the HHE and originally focused on mathematical issues 
\cite{colli1993, friedman1989, showalter1987} and the correct form of boundary conditions
\cite{glass1991, greenberg1987}. Solomon \etal\cite{solomon1985} developed an enthalpy 
formulation of the hyperbolic Stefan problem and used numerical simulations to show that increasing
the relaxation time can alter the solidification kinetics.   
Liu \etal\cite{liu2009} compared the parabolic (classical) and hyperbolic Stefan problems
in the context of thermal spray particles and concluded that flux relaxation
only influences the early stages of solidification, % where the front velocity is the greatest, 
which agrees with the earlier work by Sadd and Didlake \cite{sadd1977}. As shown by Mullis
\cite{mullis1997}, hyperbolic heat transport can strongly influence the formation of dendrites
in rapidly solidifying metal baths. Recently, the hyperbolic Stefan model has been applied to
solidification problems arising in pulsed-laser surface treatment \cite{wang2000},
cryopreservation of skin \cite{deng2003} and other biological tissues \cite{ahmadikia2012},
and cryosurgery of lung cancer \cite{kumar2017}. Sobolev \cite{sobolev1995} derived the GK equation
from a two-temperature model and coupled it to the Stefan condition to study ultra-fast melting and
solidification in the context of pulsed-laser experiments. This study, however, was restricted to the case of 
constant interface velocities and travelling-wave solutions for the temperature and flux.

In this paper, we carry out a detailed investigation of nanoscale solidification by coupling the GK equation
to the Stefan condition. Matched asymptotic expansions are used to solve the free boundary problem
without prior assumptions about the form of the solution and interface velocity. 
The systematic asymptotic analysis clearly
elucidates the relationship between non-Fourier heat transport and the kinetics of solidification, and demonstrates
the occurrence of large deviations from the classical behaviour predicted by Fourier's law. 
To the best of our knowledge, this is the first time that matched asymptotic expansions have been used to
study non-Fourier Stefan problems.

The paper is organised as follows. In Sec.~\ref{sec:model}, a one-phase model for one-dimensional
nanoscale solidification is presented. The model focuses on heat conduction through the solid as
described by the GK equation. Asymptotic 
solutions to the one-phase model are computed in Sec.~\ref{sec:asy} and used to understand how 
non-Fourier heat transport affects the solidification process. The paper concludes in 
Sec.~\ref{sec:conc}.

\section{Model formulation}
\label{sec:model}

\begin{figure}
  \centering
  \includegraphics[width=0.7\textwidth]{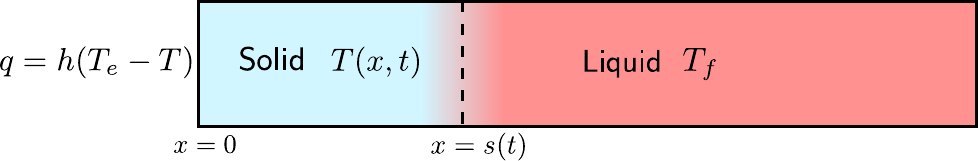}
  \caption{The solidification of a semi-infinite liquid bath that is in contact
    with a cold environment with temperature $T_\text{e}$. 
    The transfer of heat from bath into the environment drives the solidification
    process and is modelled using a Newton boundary condition with a heat
    transfer coefficient $h$. The position of the planar solid-liquid interface
    is denoted by $x = s(t)$.}
  \label{fig:bar}
\end{figure}

We consider the growth of a nanoscale solid into a semi-infinite liquid bath,
as depicted in Fig.~\ref{fig:bar}.
% Due to the inclusion of relaxation effects
% in the model, an initial condition
% for the thermal flux also needs to be prescribed.  The initial
% thermal flux is assumed to be zero, which is consistent with 
% classical models based on Fourier's law of heat conduction. 
We will assume that one side
of the bath is exposed to a cold environment that is held
at a temperature $T_\text{e}$ that is below the freezing temperature $T_\text{f}$.  The
solidification process will, therefore, be solely driven by the transfer
of heat from the bath into the environment.  Newton's law
will be used to model the exchange of thermal energy between
the solid and surrounding environment. 
The model will consist of a conservation
equations for thermal energy, the GK equation for the thermal flux, 
and appropriate boundary and initial conditions. 

We will restrict our attention to one-dimensional dynamics and thus only
consider planar solid-liquid interfaces.  
Furthermore, the initial temperature of the liquid bath is taken to be equal its freezing
temperature $T_\text{f}$. As discussed below, this will allow us to study a one-phase model.
Although this is a highly idealised situation, the simplicity of this configuration
will facilitate the mathematical analysis and provide greater insights
into the roles of nanoscale physics in solidification processes. 
The governing equations are written in terms of a
Cartesian coordinate $x$ describing the distance between the exposed solid surface
and material points in the solid and bath, and a time variable $t$.
The position of the solid-liquid interface, or solidification front, will be
denoted by $s(t)$. 

In this one-dimensional Cartesian setting, the temperature of the liquid will remain spatially
uniform and equal to $T_f$, and the thermal flux through the liquid will be zero. Thus, the
liquid will not play a role in the solidification process. Consequently,
it is sufficient to consider a one-phase model involving only the solid. 
If the initial temperature of the bath is greater than the freezing temperature, then a two-phase model
is required to capture the transfer of thermal energy from the liquid into the solid, which is expected to
reduce the rate of solidification.

\subsection{Bulk equations}

Conservation of thermal energy provides an evolution equation for the solid temperature $T$ given by 
\begin{align}
\rho c_{p} \pd{T}{t} + \pd{q}{x} = 0,
\label{dim:T}
\end{align}
where $\rho$ and $c_p$ are the density and specific heat at constant
pressure, respectively.
The thermal flux of the solid, $q$, evolves according to the one-dimensional
GK equation, 
\begin{align}
\tau_{r}\pd{q}{t} + q + k\pd{T}{x}
  = 3l^2\pdd{q}{x},
\label{dim:q_s}
\end{align}
where $\tau_r$ is the relaxation time, $k$ is the bulk
thermal conductivity, and $l$ is the phonon mean free path (MFP). 
The first term on the left-hand side of the GK equation \eqref{dim:q_s} captures memory effects
and the delayed response of the flux to a change in thermal environment. The right-hand side of
the GK equation captures non-local effects due to phonon collisions. 
Taking $\tau_r \to 0$ and $l \to 0$ or,
equivalently, $t \gg \tau_r$ and $x \gg l$, in \eqref{dim:q_s}
recovers Fourier's law, $q = -k \pdf{T}{x}$. The various limits of the GK equation
and its relationship to more general transport models are discussed in 
Sobolev \cite{sobolev1997}.

% Nanoscale heat conduction in liquids is not as well understood as for solids. However,
% phonon-based theories of liquid thermodynamics are in good agreement with experimental data 
% \cite{bolmatov2012}. Therefore, we assume that the thermal flux of the liquid, $q^\text{l}$, also
% evolves according to the GK equation
% \begin{align}
%   \tau_r^{\text{l}} \pd{q^\text{l}}{t} + q^\text{l} + k^\text{l} \pd{T^\text{l}}{x} = 3 (l^\text{l})^2
%   \pdd{q^\text{l}}{x}.
%   \label{dim:q_l}
% \end{align}

In principle, all of the physical parameters in the bulk equations
\eqref{dim:T}--\eqref{dim:q_s} can depend on
the temperature and possibly the temperature gradient \cite{jou2010};
however, they are assumed to be constant.  
The GK equation \eqref{dim:q_s} bears a marked resemblance
to the one-dimensional Navier-Stokes equation (with a linear drag term).
The temperature and flux 
play the respective roles of the pressure and fluid velocity, and the phonon MFP acts as a 
`thermal viscosity', providing resistance to flux gradients.

\subsection{Boundary conditions}

To isolate the impact of non-Fourier heat conduction on the solidification kinetics, we use
classical forms of the boundary conditions.
The Newton condition at $x = 0$, describing the exchange 
of thermal energy between the cold environment and the solid, is given by
\begin{align}
  q = h (T_\text{e} - T), \quad x = 0, \label{dim:Newton}
\end{align}
where $h$ is a heat transfer coefficient. The temperature at the interface
is taken to be equal to the freezing temperature,
\begin{align}
  T = T_f, \quad x = s(t). \label{dim:T_I}
\end{align}
Finally, a balance of thermal energy across the solid-liquid interface
yields the Stefan condition given by
\begin{align}
\rho L_m \td{s}{t} = - q, \quad x = s(t),
\label{dim:s}
\end{align}
where $L_m$ is the latent heat of solidification. 

Non-classical boundary conditions would account for the finite time that is required for the temperature and
flux to respond to a change in thermal environment \cite{sobolev1991}. In terms of the Newton condition
\eqref{dim:Newton}, accounting for this delay amounts to adding $\tau_r \d q / \d t$ to the left-hand side. 
Similarly, 
relaxation effects can lead to overheating \cite{sobolev1995, sobolev1996}, 
whereby the temperature at the interface exceeds the
equilibrium solidification temperature $T_f$, and can lead to modifications of the Stefan condition
as well \cite{sobolev1991}. We leave the study of non-classical boundary conditions as an area of future work.

\subsection{Initial conditions}

It is assumed that solidification begins from a small seed crystal of width $s_c$ that
has formed, for example, by heterogeneous nucleation, 
at the wall of the bath. We take the initial temperature 
of the seed crystal to be equal to the freezing temperature and the thermal flux
throughout the crystal to be zero. Thus,
\begin{align}
  T(x,0)  = T_f, \quad 
  q(x,0) = 0, \quad
  s(0) = s_c. \label{dim:ic}
\end{align}

A key distinction between classical and non-classical solidification problems is the role of the
seed crystal. Classical problems capture the macroscopic dynamics on length scales that
are typically much larger than those of the seed crystal and its size can be taken to be zero. 
%The corresponding heat equation for the solid phase is solved on a domain that initially
%does not exist. Remarkably, the problem for the solid remains well defined and a small-time analysis
%can be used to determine the solution for arbitrarily small crystal sizes \cite{font2015, ribera2016}.
%As shown in Appendix \ref{app:st_GK}, repeating such an analysis for the GK model leads to an 
%undetermined constant of integration. The introduction of a seed crystal enables this constant to
%be determined through the imposition of initial conditions. 
The aim of non-classical models is to describe heat transfer on small length scales that are
comparable to solid nuclei and, consequently, non-classical problems become dependent on the size of
the seed crystal.

\subsection{Non-dimensionalisation}
The problem is non-dimensionalised by defining a temperature scale $\Delta T = T_\text{f} - T_\text{e} > 0$. 
The thermal flux is driven by the temperature difference between the fluid and the adjacent cold environment.
This motivates choosing a scale for the flux by balancing terms in the Newton boundary condition
\eqref{dim:Newton}, leading to $q \sim h \Delta T$. This choice allows the decrease in temperature near the 
environment to be captured, regardless of the size of the heat transfer coefficient $h$. 
There are multiple choices for the length scale, including
the size of the seed crystal $s_c$. However, given that Fourier's law will eventually be recovered on 
length scales exceeding that of the seed crystal, we balance $q$ and $k\, \pdf{T}{x}$ in the GK equation
\eqref{dim:q_s}, yielding a length scale of $x \sim k / h$. This choice will facilitate comparing the
classical and non-classical models.
Finally, a time scale
is chosen by balancing terms in the energy equation \eqref{dim:T},
giving $t \sim \rho c_p k / h^2$. Upon writing
$x = (k / h)x'$, $t = (\rho c_p k / h^2) t'$, 
$T = T_\text{f} + (\Delta T) T'$, 
and $q = (h \Delta T) q'$
in the governing equations \eqref{dim:T}--\eqref{dim:ic},
the non-dimensional equations (upon dropping the prime) are given by
\subeq{
  \label{eqn:full}
  \begin{align}
    \pd{T}{t} + \pd{q}{x} &= 0, \label{eqn:T} \\
    \gamma \pd{q}{t} + q + \pd{T}{x} &= 
    \ell^2 \pdd{q}{x}, \label{eqn:gk}
  \end{align}
with boundary conditions
\begin{alignat}{2}
  q &= -(1 + T), &\quad x &= 0, \\
  T &= 0, &\quad x &= s(t).
\end{alignat}
The Stefan condition is
\begin{align}
  \beta \td{s}{t} = -q, \quad x = s(t),
  \label{eqn:s}
\end{align}
and the dimensionless initial conditions are
\begin{align}
  T = 0, \quad q = 0, \quad s = \varepsilon; \quad t = 0. \label{eqn:ic}
\end{align}
}
% \subeq{
% \begin{alignat}{2}
%   T &= 0, &\quad t &= 0, \\
%   q &= 0, &\quad t &= 0, \\
%   s &= \varepsilon, &\quad t &= 0.
% \end{alignat}
% }
The four dimensionless parameters that appear in the 
governing equations are defined by
\begin{align}
  \beta = \frac{L_m}{c_p \Delta T}, \quad
  \gamma = \frac{\tau_r h^2}{\rho c_p k}, \quad
  \ell = \frac{3^{1/2} l h}{k}, \quad
  \varepsilon = \frac{s_c h}{k},
\end{align}
corresponding to the Stefan number, normalised
relaxation time, MFP, and seed crystal size, respectively. 

\subsection{Model reformulations}

In some cases, the asymptotic analysis proceeds more straightforwardly
if the GK equation \eqref{eqn:gk} is re-written as
\begin{align}
\gamma \pd{q}{t} + q + \pd{T}{x} = -\ell^2\pd{}{t}\left(\pd{T}{x}\right),
\label{eqn:gk2}
\end{align}
which is obtained by first differentiating the energy equation
\eqref{eqn:T} with respect to $x$ and using the result to
eliminate the $\partial^2 q / \partial x^2$ term in \eqref{eqn:gk}.  
In the limit of zero relaxation time, $\gamma \to 0$, the GK equation \eqref{eqn:gk2} becomes
\begin{align}
q = -\pd{T}{x} -\ell^2\pd{}{t}\left(\pd{T}{x}\right),
\label{eqn:gk3}
\end{align}
indicating that non-local effects can drive, or inhibit, the thermal flux through
temporal changes in the temperature gradient.
By decomposing the flux $q$ into its
Fourier component $-\pdf{T}{x}$ and a non-classical
component $j$, so that $q = -\pdf{T}{x} + j$, the bulk equations can be
written as
\subeq{
  \label{eqn:q_decomp}
\begin{align}
  \pd{T}{t} - \pdd{T}{x} &= -\pd{j}{x}, \\
  \pd{j}{t} + \gamma^{-1} j &= (1 - \ell^2 / \gamma)\pd{}{t}\left(\pd{T}{x}\right).
\end{align}
}
% Furthermore, the energy equation \eqref{eqn:T} and the GK equation 
% \eqref{eqn:gk2} can be combined to yield an evolution equation
% for the temperature given by
% \begin{align}
% \gamma \pd{}{t}\left(\pd{T}{t} - \frac{\ell^2}{\gamma}\pdd{T}{x}\right)
% + \left(\pd{T}{t} - \pdd{T}{x}\right) = 0. \label{eqn:T2}
% \end{align}
The initial conditions \eqref{eqn:ic} imply that $j(x,0) = 0$.
From \eqref{eqn:q_decomp}, it is straightforward to see that $j$ satisfies an ordinary differential
equation in time; therefore, only two boundary conditions for this problem are required,
either for the temperature, flux, or a combination of these two quantities.  If the 
dimensionless relaxation time and MFP are such that $\ell^2 / \gamma = 1$, 
then solutions of the classical heat equation, based on Fourier's law,
are solutions to the non-classical model \eqref{eqn:q_decomp}, implying that 
non-classical effects do not lead to any distinguishing behaviour in this case.
Consequently, setting $\ell^2 / \gamma = 1$ in the asymptotic solutions will enable the
Fourier solutions to be easily recovered. 
It has been shown that if $\ell^2 / \gamma > 1$, then the temperature is dominated by diffusion;
if $\ell^2 / \gamma < 1$, then the temperature propagates as a wave 
\cite{kovacs2015, moosaie2008, van2017}.

\subsection{Numerical method}

To validate the asymptotic analysis, the full non-dimensional model \eqref{eqn:full} is
numerically solved using a semi-implicit finite difference method. The temperature $T$ and flux $q$ are treated
implicitly while the position $s$ and speed $\d s/\d t$ of the solidification front are treated explicitly; 
see, for example, Font \etal\cite{font2015} for further details. This approach effectively decouples the 
solidification and transport processes, allowing $T$ and $q$ to be (implicitly) updated for known $s$ and
$\d s / \d t$, and then $s$ and $\d s / \d t$ to be (explicitly) updated with the new flux $q$. The change of
variable $\eta = x / s(t)$ is used to transform the growing domain $0 \leq x \leq s(t)$ to a fixed domain
$0 \leq \eta \leq 1$. 
Second-order finite difference formulae are then used to discretise all spatial derivatives; central 
differences are used for interior grid points with forwards and backwards differences being employed when necessary
for boundary grid points.

\subsection{Parameter estimation}
\label{sec:params}

Tin is commonly used in theoretical and experimental studies of 
nanoscale heat transfer and phase change \cite{lai1996, ribera2016, wronski1967}. As a solid,
tin has a density of $\rho = 7180$ kg/m$^3$, a bulk thermal conductivity of
$k = 67$ W/(m$\cdot$K), and specific heat of $c_p = 230$ J/(kg$\cdot$K).
Tin solidifies at $T_\text{f} = 505$ K and has a latent heat of $L_m = 58500$ J/kg. 
The Stefan number can be parameterised in terms of the temperature difference 
$\Delta T = T_\text{f} - T_\text{e}$ as $\beta \simeq 254\text{ K}/\Delta T$ and may be treated as
large  for moderate values of $\Delta T$ up to 50 K. 
Similar conclusions are reached for other solidifying metals such as aluminium and
nickel \cite{liu2009, mullis1997}. 
To facilitate the analysis, we restrict our attention to the case of large Stefan numbers. 
The heat transfer coefficient $h$ depends on the nature of the surrounding environment and can be
difficult to estimate. Therefore, we follow Ribera and Myers \cite{ribera2016} and take
$h = 4.7\times 10^{9}$ W/(m$^2\cdot$K), which provides the closest approximation to the
fixed-temperature boundary condition that is thermodynamically reasonable \cite{jou2010}. 
Estimates of the phonon MFP range from
1 nm to 100 nm \cite{cahill2003}, with 40 nm being the `textbook' value \cite{johnson2013}. 
There is a similarly broad range of estimates for the relaxation time, which are typically 
on the order of $10^{-12}$ s to $10^{-10}$ s \cite{galenko1999, liu2009}. This leads to dimensionless 
relaxation times ranging from $\gamma = 0.2$ to $\gamma = 20$ and 
dimensionless MFPs of $\ell = 0.12$ to $\ell = 120$. The dimensionless seed crystal size
is more difficult
to estimate; however, given that $k^\text{s}/h$ defines a length scale of 14 nm, we assume that
$\varepsilon$ is small.

%============================================

%============================================

\section{Asymptotic analysis}
\label{sec:asy}

The presumed largeness of the Stefan number and the wide range of values for the 
dimensionless relaxation time and MFP can be exploited to construct asymptotic approximations to the
full model \eqref{eqn:full}.
The limit of large Stefan number corresponds to the time scale of thermal diffusion being
much smaller than the time scale of interface motion. This is typical for many materials and physical scenarios.
Consequently, it is possibly the most common approach for simplifying theoretical models 
of phase change. The relatively fast thermal diffusion rate implies that, on the time scale of interface motion, 
the temperature
can be well approximated by its quasi-steady profile.  In many cases, an analytical solution for the
quasi-steady temperature profile can be obtained,
reducing the Stefan problem to an ordinary differential equation for the 
position of the free boundary; see, for example, Myers \etal\cite{font2015, ribera2016, florio2016}.
In Sec.~\ref{sec:O1}, we consider the case when 
non-dimensional relaxation time, $\gamma$, and MFP, $\ell$, are order one or smaller. The solidification
kinetics are well described by the classical solution, indicating that non-Fourier heat conduction
does not play a large role in this parameter regime. 
The cases when $\ell \gg 1$ and $\gamma \gg 1$ are considered in Sec.~\ref{sec:large_bl} and
Sec.~\ref{sec:large_bg}, respectively. Substantial deviations from the classical solidification kinetics
are found in these cases.

\subsection{Order-one relaxation time and MFP}
\label{sec:O1}

We consider the asymptotic limit as $\beta \to \infty$ with $\varepsilon \ll \beta^{-1}$.
There are four key time regimes to consider. The first, given by $t = O(\varepsilon^2)$, 
describes the initial cooling of the seed crystal and it 
captures the transient development of the temperature gradient and thermal flux. The dynamics in this
regime are dominated by non-classical effects. Remarkably, however, the governing equations reduce to
Fourier's law with an effective thermal conductivity that depends
on the dimensionless relaxation time and
MFP. The second time regime, $t = O(\varepsilon \beta)$, captures the initial growth of the solid. The
temperature has a quasi-steady profile with a constant gradient (in time and space)
that differs from the classical case.
In the third regime, $t = O(1)$, non-local effects drive the temperature and flux to their
classical profiles. In essence, the third regime captures the transition from non-classical to
classical heat conduction. The size of the solid remains relatively unchanged up to this point.
Finally, the fourth time regime, $t = O(\beta)$, captures the quasi-steady growth of the solid.
Non-classical effects are small on this large time scale and the classical quasi-steady Stefan problem
is recovered at leading order. 

\emph{First time regime}: The variables are rescaled in order to capture the evolution of the
temperature and flux from their initial conditions on a length scale that coincides with the
initial size of the solid. Thus, we choose a length scale of $O(\varepsilon)$ and rescale
time, temperature, and flux to ensure that the time derivatives of the latter two quantities enter
the leading-order problem. This leads to a rescaling given by
$t = \varepsilon^2 \tilde{t}$, $x = \varepsilon \tilde{x}$, $s = \varepsilon \tilde{s}$,
$T = \varepsilon \tilde{T}$, and $q = \tilde{q}$. The leading-order bulk equations are
\subeq{
\label{pm:r0}
\begin{align}
  \pd{\tilde{T}}{\tilde{t}} + \pd{\tilde{q}}{\tilde{x}} &= 0, \label{pm:r0_energy}\\
  \gamma \pd{\tilde{q}}{\tilde{t}} &= -\ell^2 \pd{}{\tilde{t}}\left(\pd{\tilde{T}}{\tilde{x}}\right) 
\label{pm:r0_GK},
 \end{align}
}
subject to $\tilde{T} = \tilde{q} = 0$ at $\tilde{t} = 0$. 
Upon integration of \eqref{pm:r0_GK}
in time and imposing the initial conditions, we find that 
$\tilde{q} = -(\ell^2/\gamma) \pdf{\tilde{T}}{\tilde{x}}$,
recovering Fourier's law with an effective thermal
conductivity $\ell^2 / \gamma$ that depends on non-classical parameters. 
The leading-order Stefan condition is given by $\d \tilde{s} / \d \tilde{t} = 0$, 
with $\tilde{s}(0) = 1$,
indicating that crystal growth is negligible on this time scale.
The leading-order problem is simply a classical heat
conduction problem on a fixed domain:
\subeq{
\label{pm:r0_thermal}
\begin{align}
  \pd{\tilde{T}}{\tilde{t}} = \frac{\ell^2}{\gamma} \pdd{\tilde{T}}{\tilde{x}},
\end{align}
subject to
\begin{alignat}{2}
  \frac{\ell^2}{\gamma} \pd{\tilde{T}}{\tilde{x}} &= 1, &\quad \tilde{x} &= 0, \\
  \tilde{T} &= 0, &\quad \tilde{x} &= 1, \\
  \tilde{T} &= 0, &\quad \tilde{t} &= 0.
\end{alignat}
}
The large-time limit of the solution to \eqref{pm:r0_thermal}, the flux, and the position
of the solidification front, written in terms of the original 
dimensionless variables, is given by
\begin{align}
  T \sim -(\gamma / \ell^2) (\varepsilon - x), \quad q \sim -1, \quad s \sim \varepsilon,
  \label{pm:r0_match}
\end{align}
which will be used to match into the next time regime.

\emph{Second time regime}:
The second time regime is defined by $t = O(\varepsilon \beta)$, which arises from
balancing terms in the Stefan condition
given that $q = O(1)$ and $s = O(\varepsilon)$ from the matching
condition \eqref{pm:r0_match}. Furthermore, the matching condition
for the temperature implies that $T = O(\varepsilon)$ (recall that $\gamma / \ell^2 = O(1)$). 
The variables are therefore rescaled according 
to $x = \varepsilon \bar{x}$, $s = \varepsilon \bar{s}$,
and $T = \varepsilon \bar{T}$. We also write $q = -1 + \varepsilon \bar{q}$
to ensure the matching condition is satisfied
and to balance terms in the Newton boundary condition.
The rescaled bulk equations become
\subeq{
\begin{align}
  \pd{\bar{q}}{\bar{x}} &= O(\beta^{-1}), \\
  \pd{}{\bar{t}}\left(\pd{\bar{T}}{\bar{x}}\right) &= O(\varepsilon \beta),
                                                     \label{bar:r2_GK}
\end{align}
}
and the boundary conditions are $\bar{q} = -\bar{T}$ at $\bar{x} = 0$ and $\bar{T} = 0$ at
$\bar{x} = \bar{s}(\bar{t})$.  
%\subeq{
%\label{bar:r2}
%\begin{align}
%  \beta^{-1} \pd{T}{t} + \pd{Q}{x} &= 0, \label{bar:r2_energy} \\
%  \beta^{-1}\gamma \pd{Q}{t} + 1 + \varepsilon Q
%  + \pd{T}{x} &= -\varepsilon^{-1} \beta^{-1} \ell^2\pd{}{t}\left(\pd{T}{x}\right), \label{bar:r2_GK}
%\end{align}
%subject to the following boundary conditions:
%\begin{alignat}{2}
%Q &= -T, &\quad x &= 0, \\
%T &= 0, &\quad x &= s(t).
%\end{alignat}
%The rescaled Stefan condition is
%\begin{align}
%  \td{s}{t} = 1 + \varepsilon Q, \quad x = s(t).
%\end{align}
%}
The relevant matching conditions are given by
$\bar{T} \sim -(\gamma / \ell^2) (1 - \bar{x})$ and $\bar{s} \sim 1$
as $\bar{t} \sim 0$.  
Solving the leading-order problem yields the solution
\subeq{
  \label{bar:r2_soln}
  \begin{align}
    \bar{T}(\bar{x},\bar{t}) &= -(\gamma / \ell^2) (\bar{s}(\bar{t}) - \bar{x}), \label{bar:r2_T}\\
    \bar{q}(\bar{x},\bar{t}) &= (\gamma / \ell^2) \bar{s}(\bar{t}), \\
    \bar{s}(\bar{t}) &= 1 + \bar{t}. \label{bar:r2_s}
  \end{align}
}
For $\bar{t} \gg 1$, the leading-order solution \eqref{bar:r2_soln}
can be written in terms of the original non-dimensional variables as
$T \sim -\beta^{-1} (\gamma / \ell^2) (t - \beta x)$,
$q \sim -1 + \beta^{-1} (\gamma / \ell^2) t$,
and $s \sim \beta^{-1} t$.

\emph{Third time regime}:
The scales for the third regime, $t = O(1)$, arise from the $\bar{t} \gg 1$ limit of the
solution in the second time regime and, in essence, promote the $O(\varepsilon \beta)$ terms in
\eqref{bar:r2_GK} to leading order. The time scale of $t = O(1)$ is determined by balancing terms in
the GK equation \eqref{eqn:gk} given that $s \sim \beta^{-1} t$, $T \sim \beta^{-1}t$, and
$q \sim -1 + \beta^{-1} t$. The variables are then rescaled as
$t = \hat{t}$, $x = \beta^{-1} \hat{x}$, $s = \beta^{-1} \hat{s}$,
$T = \beta^{-1} \hat{T}$, and $q = -1 + \beta^{-1} \hat{q}$. 
The energy equation is given by 
$\pdf{\hat{q}}{\hat{x}} = O(\beta^{-1})$.
Similarly, the rescaled GK equation reads 
$-1 + \pdf{\hat{T}}{\hat{x}} = -\ell^2\pdf{(\pdf{\hat{T}}{\hat{x}})}{\hat{t}} + O(\beta^{-1})$,
corresponding to an extension of Fourier's law (the left-hand side) 
that includes non-local effects (the right-hand side). 
The boundary conditions are $\hat{q}(0,\hat{t}) = -\hat{T}(0,\hat{t})$ and 
$\hat{T}(\hat{s}(\hat{t}),\hat{t}) = 0$,
and the Stefan condition is $\d \hat{s} / \d \hat{t} = 1 + O(\beta^{-1})$ %- \beta^{-1}\hat{q}(\hat{s}(\hat{t}),\hat{t})$. 
The relevant matching conditions are 
$\hat{T} \sim -(\gamma / \ell^2)(\hat{t} - \hat{x})$, 
$\hat{q} \sim (\gamma / \ell^2)\hat{t}$, and
$\hat{s} \sim \hat{t}$ as $\hat{t} \sim 0$. 
The leading-order solutions for the temperature, the flux, and %the first two terms in the solution for 
the position of the solidification front are
\subeq{
  \label{bar:r3}
  \begin{align}
    \hat{T}(\hat{x},\hat{t}) &= -(\hat{s}(\hat{t}) - \hat{x})\left[1 - (1 - \gamma / \ell^2)\e^{-\hat{t}/\ell^2}\right] \label{bar:r3_T}\\
    \hat{q}(\hat{x},\hat{t}) &= \hat{s}(\hat{t})\left[1 - (1 - \gamma / \ell^2)\e^{-\hat{t}/\ell^2}\right], \\
    \hat{s}(\hat{t}) &= \hat{t} \label{bar:r3_s}
    % + \beta^{-1}\left[(\ell^2 - \gamma)\left(\ell^2 - (\hat{t} + \ell^2)\e^{-\hat{t}/\ell^2}\right)
                       %- \hat{t}^2 / 2\right].
\end{align}
}
The classical solution can be obtained by taking the limit 
$\gamma / \ell^2 \to 1$, which shows that pure Fourier conduction leads to a
temperature profile with a constant gradient and a flux $q \sim -1 + t$ that decreases (in magnitude)
linearly in time. The role of non-classical effects is to introduce a time dependence that
drives the temperature and flux to their classical profiles.

\emph{Fourth time regime}: The fourth time regime captures the growth of the solid over
long time scales. Balancing terms in the Stefan condition with $s = O(1)$ and $q = O(1)$
requires $t = O(\beta)$. Thus, time is written as $t = \beta \tau$ and the other variables are
unchanged. The rescaled problem is given by
\subeq{
  \label{bar:r4}
  \begin{align}
    \beta^{-1} \pd{T}{\tau} + \pd{q}{x} &= 0, \label{bar:r4_energy} \\
    \beta^{-1} \gamma \pd{q}{t} + q + \pd{T}{x} &= 
    \ell^2 \pdd{q}{x} \label{bar:r4_GK},
  \end{align}
with boundary conditions
\begin{alignat}{2}
  q &= -(1 + T), &\quad x &= 0, \\
  T &= 0, &\quad x &= s(\tau).
\end{alignat}
The Stefan condition is
\begin{align}
  \td{s}{\tau} = -q, \quad x = s(\tau),
\end{align}
}
with matching conditions in time that will be specified below. Capturing the non-classical
effects on this time scale requires the calculation of higher-order terms in the asymptotic expansions
of the variables. Therefore, the solution to \eqref{bar:r4} is sought as an 
asymptotic expansion of the form
$T = T_0 + \beta^{-1} T_1 + O(\beta^{-2})$,
$q = q_0 + \beta^{-1} q_1 + O(\beta^{-2})$, and
$s = s_0 + \beta^{-1} s_1 + O(\beta^{-2})$.

The leading-order contribution to the
energy equation \eqref{bar:r4_energy} 
shows that the flux is constant in space
and given by $q_0(x,\tau) \equiv -(1 + T_0(0,\tau))$,
which implies that the non-local terms drop out of the
GK equation \eqref{bar:r4_GK}. 
Moreover, due to the assumption that $\gamma = O(1)$, flux
relaxation is
also a higher-order effect. The leading-order GK equation reduces to Fourier's
law, $q_0 = -\pdf{T_0}{x}$. The temperature profile 
and flux can be written as
\subeq{
\label{bar:r4_0}
\begin{align}
  T_0(x,\tau) &= \frac{x - s_0(\tau)}{1 + s_0(\tau)}, \label{bar:r4_0_T}\\
  q_0(x,\tau) &= -\frac{1}{s_0(\tau) + 1}.
\end{align}
Integrating the Stefan condition and using the matching condition
$s_0(0) = 0$ yields
\begin{align}
s_0(\tau) = -1 + \sqrt{1 + 2 \tau}. \label{bar:r4_s0}
\end{align}
}
The leading-order solutions for the position of the solidification front
\eqref{bar:r2_s}, \eqref{bar:r3_s}, and \eqref{bar:r4_s0} can be combined
into a composite solution \cite{bender2013, hinch1991} given by
\begin{align}
  s(t) = \varepsilon + \sqrt{1 + 2 \beta^{-1} t} - 1, \label{bar:comp_s}
\end{align}
which is valid for all times. 

The next-order problem contains contributions from non-classical
terms and can be written as
\subeq{
\begin{align}
  \pd{T_0}{\tau} + \pd{q_1}{x} &= 0, \\
  \gamma \td{q_0}{\tau} + q_1 + \pd{T_1}{x} &= \ell^2\pdd{q_1}{x}, \label{bar:r4_GK_2}
\end{align}
with boundary conditions 
\begin{alignat}{2}
  q_1 &= -T_1, &\quad x &= 0, \\
  T_1 &= -\pd{T_0}{x} s_1, &\quad x &= s_0(\tau).
\end{alignat}
The Stefan condition is given by
\begin{align}
  \td{s_1}{\tau} = -q_1, \quad x = s_0(\tau).
\end{align}
}
Matching to the solution in the first time regime gives
that $s_1(0) = 0$. Before attempting to solve the problem, it is illustrative to
first rewrite the GK equation \eqref{bar:r4_GK_2} using the solution
for $q_0$ and the fact that $\partial^2 q_1 / \partial x^2
= \d q_0 / \d \tau$, which yields
\begin{align}
  q_1 + \pd{T_1}{x} = (\gamma - \ell^2)\td{}{\tau}\left(\frac{1}{s_0(\tau) + 1}\right)
  = -\frac{\gamma - \ell^2}{(s_0(\tau) + 1)^3}.
  \label{bar:r2_q1}
\end{align}
Therefore, the flux is simply given by Fourier's law with a 
time-dependent source term, the magnitude of which is 
characterised by the grouped parameter
$\gamma - \ell^2$. 
Using \eqref{bar:r2_q1}, the flux $q_1$ can be eliminated from
the governing equations and the problem can be formulated
solely in terms of the temperature $T_1$, as in the 
classical Stefan problem. Carrying out this
process yields the system of equations given by
\subeq{
  \begin{align}
    \pdd{T_1}{x} = \pd{T_0}{\tau},
\end{align}
with boundary conditions
\begin{alignat}{2}
  \pd{T_1}{x} &= T_1 + (\gamma - \ell^2)\td{}{\tau}\left(\frac{1}{s_0(\tau) + 1}\right), 
                &\quad x &= 0, \\
  T_1 &= -\frac{s_1(\tau)}{s_0(\tau) + 1}, &\quad x &= s_0(\tau),
\end{alignat}
and the Stefan condition
\begin{align}
  \td{s_1}{\tau} = \pd{T_1}{x} - (\gamma - \ell^2)\td{}{\tau}\left(\frac{1}{s_0(\tau) + 1}\right),
  \quad x = s_0(\tau).
\end{align}
}
An analytical solution for the temperature $T_1$ can be found and used
in the Stefan condition to find that the correction to the solidification front
evolves according to
\begin{align}
  \td{s_1}{\tau} + \frac{s_1(\tau)}{(s_0(\tau) + 1)^2} = 
  -\frac{1}{3}\left[1 - \frac{1}{(s_0(\tau) + 1)^3}\right]\td{s_0}{\tau}
  + \frac{(\gamma - \ell^2)\, s_0}{(s_0(\tau) + 1)^3}\td{s_0}{\tau}.
  \label{bar:r2_s1_ode}
\end{align}
Equation \eqref{bar:r2_s1_ode} can be integrated and the solution for $s_1$ can be
written in terms of $s_0$ as
\begin{align}
  s_1(\tau) = -\frac{1}{6}\frac{s_0(\tau)^2(3 + s_0(\tau))}{(s_0(\tau) + 1)^2}
  + (\gamma - \ell^2)\left[\frac{\log(s_0(\tau) + 1)}{s_0(\tau) + 1}
  - \frac{s_0(\tau)}{(s_0(\tau) + 1)^2}\right]. \label{bar:r2_s1}
\end{align}
For small and large times, $\tau \ll 1$ and $\tau \gg 1$, respectively, we find that
\subeq{
\label{bar:r2_s1_limits}
\begin{alignat}{2}
  s_1 &\sim \frac{1}{2}\left(\gamma - \ell^2 - 1\right) \tau^2, &\quad \tau &\ll 1, \\
  s_1 &\sim -\frac{1}{6}\left((2\tau)^{1/2} + 1 - \frac{3}{(2\tau)^{1/2}}\right)
  + \frac{\gamma - \ell^2}{(2\tau)^{1/2}}\left(\frac{1}{2} \log(2\tau) - 1\right), 
    &\quad
    \tau &\gg 1.
\end{alignat}
}
Thus, for small times, the role of classical and non-classical effects
in the correction to the position of the solidification front are comparable. 
However, for large times,
non-classical effects become sub-dominant. This can be explained on physical
grounds: on large time scales, the solid can grow far beyond the phonon
MFP, reducing the influence of non-local effects, and the flux, like the temperature, 
will have relaxed to its quasi-steady state. Furthermore, both equations in
\eqref{bar:r2_s1_limits} show that non-classical effects can either facilitate
or inhibit solidification depending on the sign of $\gamma - \ell^2$. On one hand, a larger value of 
$\gamma$ corresponds to a slower relaxation (i.e., increase) 
of the flux from its initial value of $q_0 = -1$, 
thereby enhancing thermal transport from the interface to the environment and
accelerating the solidification process.  On the other hand, the combination of non-local
effects and a decreasing temperature gradient (in time)
inhibit the transport of heat throughout the solid
(see \eqref{eqn:gk3}), resulting in a slower solidification process as $\ell$ is increased.

\begin{figure}
\centering
\subfigure[]{\includegraphics[width=0.49\textwidth]{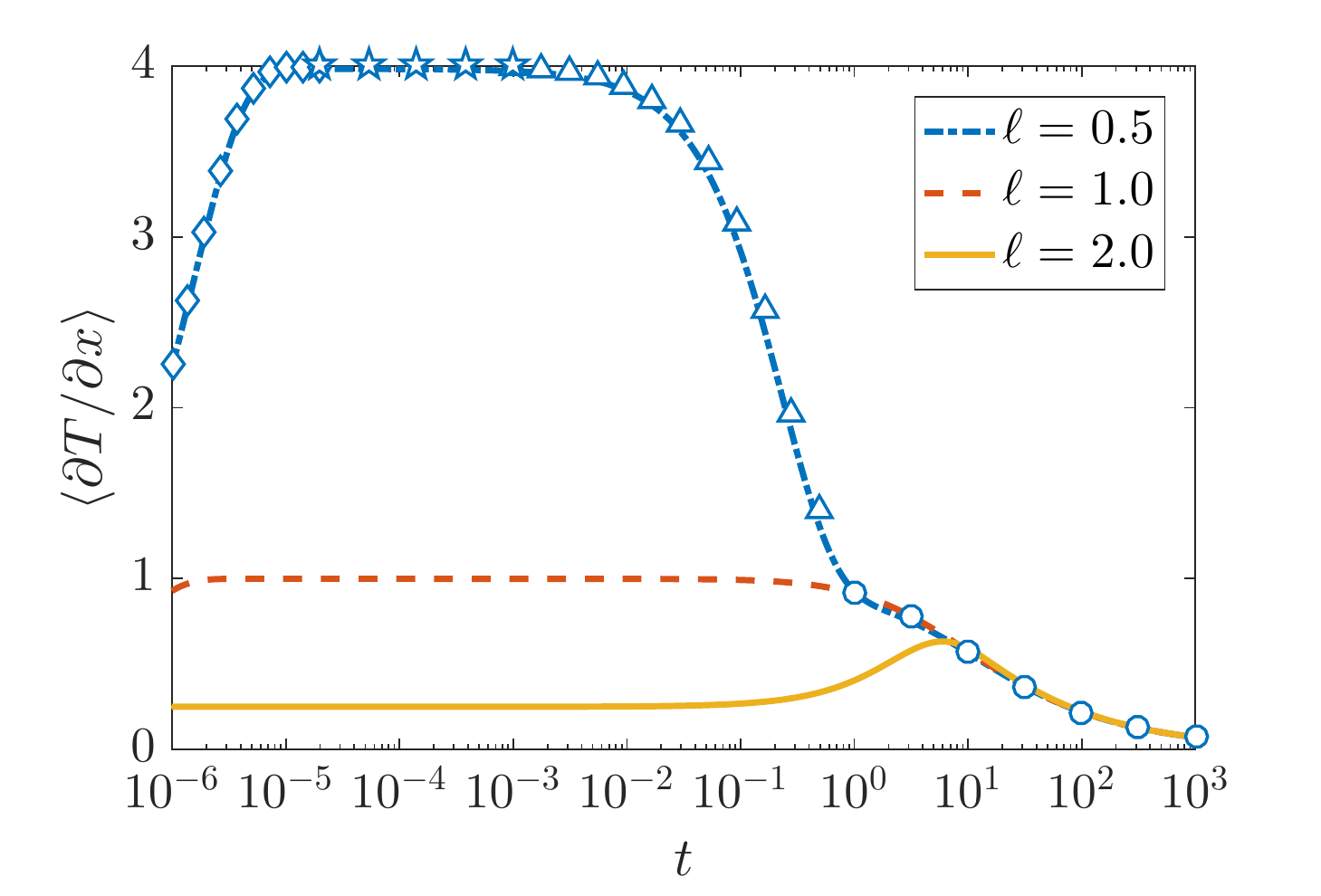}}
\subfigure[]{\includegraphics[width=0.49\textwidth]{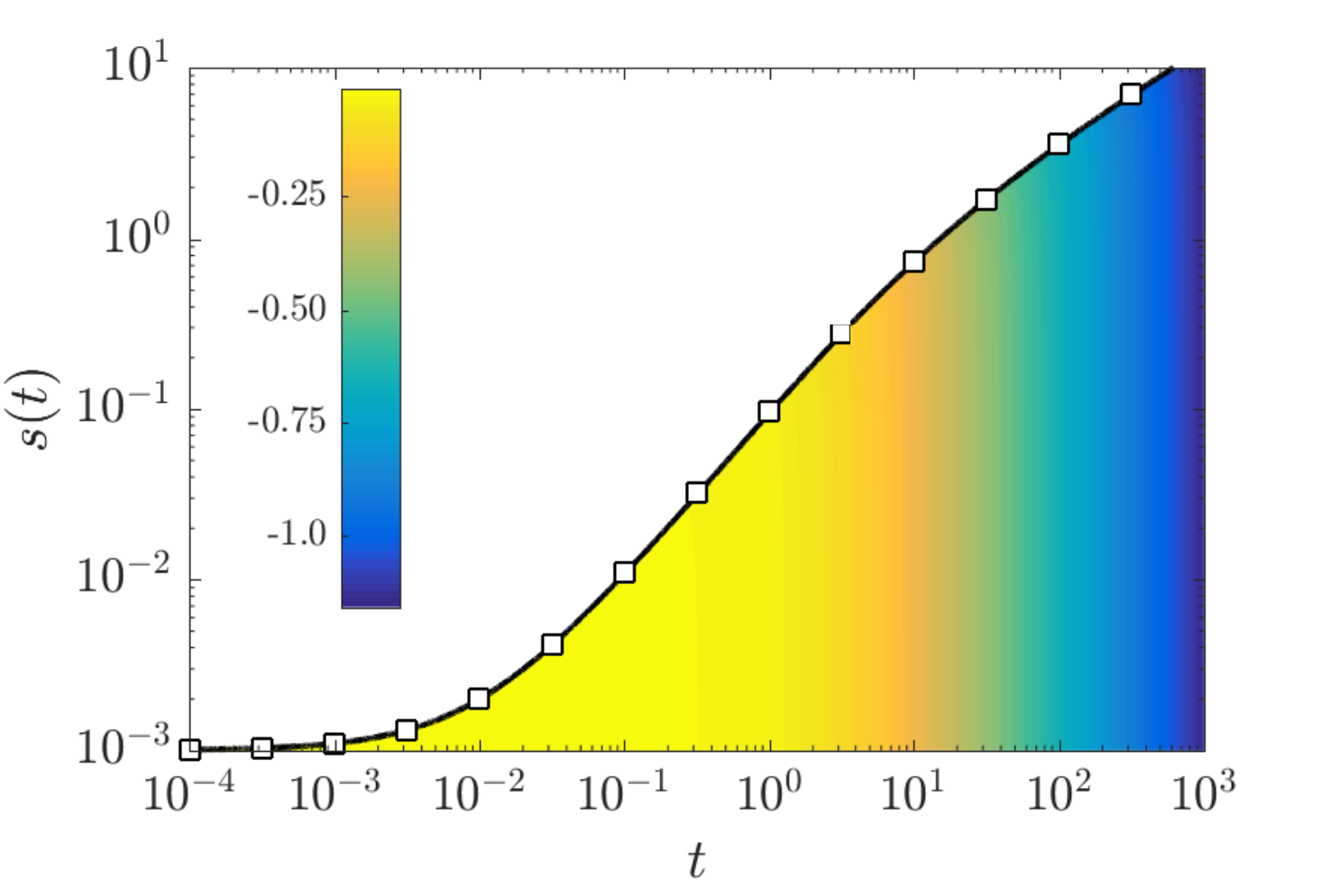}}
\caption{Solidification dynamics in the limit of large Stefan number. Left (a):
  Evolution of the mean temperature gradient $\langle \pdf{T}{x}\rangle \equiv
  [T(s(t),t) - T(0,t)] / s(t)$ for different values of $\ell$ with a fixed
  value of $\gamma = 1$.
  Right (b): Evolution of the solidification front with the logarithm of the flux,
  $\log_{10} (-q)$, superposed as a 
  colour gradient. 
  % The inset shows a comparison of a numerical solution 
  % with the correction to the position of the melt front given by
  % \eqref{bar:r2_s1}. 
  Lines and diamonds represent numerical solutions of the full model \eqref{eqn:full}
  and the reduced model in the first time regime \eqref{pm:r0_thermal}.
  Stars, triangles, circles, and squares correspond to 
  \eqref{bar:r2_T}, \eqref{bar:r3_T}, \eqref{bar:r4_0_T}, and \eqref{bar:comp_s}, respectively.
  In both panels, $\gamma = 1$, $\beta = 10$ and $\varepsilon = 10^{-3}$. A value of $\ell = 0.5$
  was used in panel (b).}
\label{fig:large_beta}
\end{figure}

Figure \ref{fig:large_beta} shows the representative dynamics that occur
in the limit of large Stefan number, using $\gamma = 1$, $\beta = 10$, and $\varepsilon = 10^{-3}$.
Figure \ref{fig:large_beta} (a) 
illustrates the evolution of the mean temperature gradient, defined by 
\begin{align}
\left<\pd{T}{x} \right> \equiv \frac{T(s(t),t) - T(0,t)}{s(t)},
\end{align}
for different values of $\ell$.
As indicated by the asymptotic analysis, the temperature quickly settles into a linear profile;
therefore, the mean temperature gradient contains all of the information about how the temperature profile
evolves in time. Since the value of $\gamma$ is fixed
at $\gamma = 1$, the values of $\ell = 0.5$, $1.0$, and $1.5$ correspond
to the three distinguishing cases: $\gamma / \ell^2 > 1$, $\gamma / \ell^2 = 1$ (Fourier),
and $\gamma / \ell^2 < 1$, respectively. Compared to the Fourier case, the temperature
gradients in the non-classical cases exhibit much richer dynamics as they evolve
through the multiple time regimes that were elucidated by the asymptotic analysis.
In the classical case, many of these regimes appear blended together because the
temperature gradient remains constant in time with a value given by
$\pdf{T}{x} = \gamma / \ell^2 = 1$. 
When $\gamma / \ell^2 > 1$ ($\gamma / \ell^2 < 1$), the temperature profile initially
has a stronger (weaker) gradient due to the reduced (enhanced) effective thermal conductivity of
the solid. Non-local effects drive the temperature gradients to their classical value and
all three curves converge when $t \simeq 10 = O(\beta)$. 
In the case of $\ell = 0.5$, the numerical
approximation of the mean temperature gradient is compared with the asymptotic solutions in
all four time regimes. Using the values of $\beta = 10$ and $\varepsilon = 10^{-3}$, these four regimes are given
by $t \simeq 10^{-6}$, $10^{-2}$, $1$, and $10$. 
Diamonds correspond to the numerical solution of \eqref{pm:r0_thermal}, valid in the first regime, which describes
the rapid initial development of the temperature gradient on $t = O(\varepsilon^2)$ time scales. 
Stars, diamonds, and circles denote the asymptotic solutions \eqref{bar:r2_T}, \eqref{bar:r3_T}, and 
\eqref{bar:r4_0_T} in the second, third, and fourth regimes, respectively. In all cases, the agreement between
asymptotics and numerics is excellent.

Figure \ref{fig:large_beta} (b) shows the evolution of the solidification front in the case when
$\ell = 0.5$. Superposed on this
panel is the logarithm of the flux, $\log_{10} (-q)$, which is uniform in space 
and decreases from $q \simeq 1$ when
$t \simeq 10^{-3}$ to $q \simeq 10^{-1}$ when $t \simeq 10^3$. The squares correspond to the
leading-order composite solution for the position of the solidification front
\eqref{bar:comp_s}, which does not contain contributions from the non-classical terms associated
with the GK equation.
The excellent agreement between the asymptotic and numerical solutions confirms that
non-classical effects do not play a significant role in the dynamics of solidification in this 
parameter regime. This is due to flux relaxation and non-local effects being negligible
over the large time scales on which appreciable growth occurs.

% By integrating the energy equation \eqref{??} over the domain, 
% and using the Newton condition \eqref{???} and Stefan conditions,
% the correction to the melt front is given by
% \begin{align}
%   s_1(\tau) = -\int_{0}^{s_0(\tau)} T_0(x,\tau)\,\d x
%   - \int_{0}^{\tau} T_1(0,\tau')\,\d \tau' + B
% \end{align}
% where $B$ is determined by matching.

% Thus, we can conclude that non-classical effects
% play a very small in the melting problem 
% when the Stefan number is large. 

%\begin{figure}
%  \centering
%  \subfigure[$\beta = 1$]{\includegraphics[width=0.45\textwidth]{figs/cart_s_beta_one.pdf}}
%  \subfigure[$\beta = 10$]{\includegraphics[width=0.45\textwidth]{figs/cart_s_beta_large.pdf}}
%  \caption{\note{Include pre-melt region.} Evolution of the solid-liquid interface in a melting rectangular bar.
%    Non-dimensional numbers are $\gamma = 1$ and $\ell = 0.5$.}
%  \label{fig:cart_s}
%\end{figure}

%==============================================

\subsection{Limit of large MFP}
\label{sec:large_bl}

We now consider the asymptotic limit as $\ell \to \infty$ and $\beta \to \infty$ 
with $\varepsilon \ll \beta^{-1} \ll 1$.
The relative sizes of $\beta$ and $\ell$ will be discussed further below. There are up to five time regimes 
to consider. The first two, defined by
$t = O(\varepsilon^2 \ell^{-2})$ and $t = O(\beta \varepsilon)$, describe the initial evolution of the
temperature and flux due to the outflow of heat into the environment, and the growth of the solid 
on length scales associated with the seed crystal. 
The third and fourth regimes, $t = O(1)$ and $t = O(\ell \beta^{1/2})$, essentially capture
the same physics, whereby the flux is driven by temporal changes in the temperature gradient, 
but slightly differ in their mathematical structure. If $\beta \ll \ell^2$, then there is a
fifth time regime, $t = O(\ell^2)$, which
captures the transition to classical quasi-static solidification, with a flux that is driven by
temperature gradients.

\emph{First and second time regimes:} The first two time regimes
are mathematically identical to those previously considered when $\ell = O(1)$. 
The limiting behaviour is 
\subeq{
  \label{large_bl:12}
  \begin{align}
    % T &\sim -\gamma \ell^{-2} \beta^{-1}(s(t) - \beta x), \\
    % q &\sim -1 + \gamma \ell^{-2} \beta^{-1} t, \\
    % s &\sim \beta^{-1} t,
    T \sim -\gamma \ell^{-2} \beta^{-1}(s(t) - \beta x), \qquad
    q \sim -1 + \gamma \ell^{-2} \beta^{-1} t, \qquad
    s \sim \beta^{-1} t,
  \end{align}
}
for $\beta \varepsilon \ll t \ll 1$.

\emph{Third time regime}: Scales for the temperature, flux, and the position of the solidification front
(and hence space) can be obtained from the limiting behaviour in the second regime \eqref{large_bl:12}. 
The time scale is selected by balancing terms in the GK equation. The only sensible balance occurs
between $q = O(1)$ and $\ell^{-2} \partial(\pdf{T}{x})/\partial t = O(t^{-1})$, which requires 
$t = O(1)$, and implies that the flux is driven by temporal changes in the temperature gradient.
 Thus, the variables are rescaled as $t = \hat{t}$, 
$x = \beta^{-1} \hat{x}$, $s = \beta^{-1} \hat{s}$, 
$T = \ell^{-2} \beta^{-1} \hat{T}$,
and we write $q = -1 + \ell^{-2} \beta^{-1} \hat{q}$. The leading-order bulk equations are
given by $\partial (\pdf{\hat{T}}{\hat{x}}) /\partial \hat{t} = 1$ and $\pdf{\hat{q}}{\hat{x}} = 0$. 
Matching conditions for the temperature and solidification front are
$\hat{T} \sim -\gamma (\hat{s} - \hat{x})$ and $\hat{s} \sim \hat{t}$ 
as $\hat{t} \sim 0$. Boundary conditions are 
$\hat{q}(0,\hat{t}) = -(1 + \hat{T}(0,\hat{t}))$ and
$\hat{T}(\hat{s}(\hat{t}),\hat{t}) = 0$. The Stefan condition is given by
$\d \hat{s} / \d \hat{t} = 1 + O(\ell^{-2}\beta^{-1})$. The leading-order solution can be written as
\subeq{
  \begin{align}
    \hat{T}(\hat{x},\hat{t}) &= -(\hat{t} + \gamma) (\hat{s}(\hat{t}) - \hat{x}), \\
    \hat{q}(\hat{x},\hat{t}) &= (\hat{t} + \gamma)\hat{s}(\hat{t}), \\
    \hat{s}(\hat{t}) &= \hat{t}, \label{large_bl_r1:s}
  \end{align}
}
which is consistent with the large-$\ell$ expansion of \eqref{bar:r3}. Equation \eqref{large_bl_r1:s}
indicates that the growth kinetics of the solid are linear when $t = O(1)$, as in the case of
classical Fourier conduction, despite the fact that the flux is driven by a non-classical 
mechanism. This is a result of the flux changing by only a very small amount in this time regime;
recall that $q = -1 + \ell^{-2}\beta^{-1}\bar{q}$. 

\emph{Fourth time regime}: The fourth time regime is an extension of the third which captures
$O(1)$ changes in the flux; thus, we take $q = O(1)$. Balancing terms in the Newton condition under
this assumption leads to $T = O(1)$. Length and time scales can be obtained by 
balancing both terms in the Stefan condition along with
$q$ and $\partial (\pdf{T}{x})/\partial t$ in the GK equation, yielding
$x = O(\ell \beta^{-1/2})$ and $t = O(\ell \beta^{1/2})$. Therefore,
the variables are rescaled as $t = \ell \beta^{1/2} \bar{t}$, 
$x = \ell \beta^{-1/2} \bar{x}$, $s = \ell \beta^{-1/2} \bar{s}$, 
$T = \bar{T}$, and $q = \bar{q}$. 
The rescaled energy and GK equations can be written as
\subeq{
\begin{align}
  \pd{\bar{q}}{\bar{x}} &= O(\beta^{-1}),     \label{large_bl_r4:energy} \\
  \ell^{-1} \beta^{-1/2} \gamma \pd{\bar{q}}{\bar{t}} + 
  \bar{q} + \alpha \pd{\bar{T}}{\bar{x}} &= -\pd{}{\bar{t}}\left(\pd{\bar{T}}{\bar{x}}\right),
    \label{large_bl_r4:gk}
\end{align}
}
where $\alpha = \ell^{-1} \beta^{1/2}$. For the moment, we will assume that
$\alpha = O(1)$. The matching conditions for this problem are given by
$\bar{T} \sim -\bar{t} (\bar{s}(\bar{t}) - \bar{x})$ and $\bar{s} \sim \bar{t}$ as $\bar{t} \sim 0$. 
Differentiation of the GK equation with respect to $\bar{x}$ yields an evolution equation
for the curvature of the temperature profile, 
\begin{align}
  \pd{}{\bar{t}}\left(\pdd{\bar{T}}{\bar{x}}\right) = -\alpha \pdd{\bar{T}}{\bar{x}}.
\end{align}
From the matching condition, $\partial^2 \bar{T} / \partial \bar{x}^2 \sim 0$
as $\bar{t} \sim 0$; therefore, the curvature of the temperature profile remains
zero for all time, implying that the solution for the temperature  can be written
as $\bar{T}(\bar{x},\bar{t}) = -G(\bar{t})(\bar{s}(\bar{t}) - \bar{x})$, 
where the temperature gradient  $G$ is to be determined. The 
energy balance \eqref{large_bl_r4:energy} gives that the flux is constant in space to leading
order and from the Newton condition we find that
$\bar{q}(\bar{x},\bar{t}) = -1 + G(\bar{t})\bar{s}(\bar{t})$. 
Using the solutions for the temperature and
the flux in the GK equation gives an evolution equation for $G$, 
which, when coupled with the Stefan condition, results in a 
two-dimensional system of nonlinear ODEs given by
\subeq{\label{eqn:Gs}
  \begin{align}
    \td{G}{\bar{t}} &= 1 - (\bar{s}(\bar{t}) + \alpha) G(\bar{t}), \\
    \td{\bar{s}}{\bar{t}} &= 1 - G(\bar{t})\bar{s}(\bar{t}).
  \end{align}
}
Initial conditions are given by $G(0) = 0$ and $\bar{s}(0) = 0$.

To understand the dynamics that occur in the fourth time regime,
it is insightful to first consider the case when $\alpha \ll 1$.
Subtraction of 
the evolution equations \eqref{eqn:Gs} followed by integration
in time shows
that $G(\bar{t}) = \bar{s}(\bar{t})$; hence, the Stefan condition becomes
$\d \bar{s} / \d \bar{t} = 1 - \bar{s}^2$, which can be solved to find that
\begin{align}
  \bar{s}(\bar{t}) = \tanh \bar{t}.
  \label{large_bl_r4:s}
\end{align}
Interestingly, the position of the solidification front reaches a steady state
given by $\bar{s}^* = 1$, signifying the end of the solidification process in this
time regime. 
%, consistent with the large-$\beta$ expansion
%of the equilibrium front position given by the solution of \eqref{eqn:s_star}. 
To capture the additional solidification that occurs,
we can introduce a fifth time regime by scaling the original
dimensionless variables as $t = \ell^2 \check{t}$, $x =\ell \beta^{-1/2} \check{x}$,
$s = \ell \beta^{-1/2} \check{s}$, $T = \check{T}$, 
and $q = \ell^{-1}\beta^{1/2} \check{q}$. The leading-order energy and GK equations are given by
$\pdf{\check{q}}{\check{x}} = 0$ and
\begin{align}
  \check{q} + \pd{\check{T}}{\check{x}} = -\pd{}{\check{t}}\left(\pd{\check{T}}{\check{x}}\right),
  \label{large_bl_r5:gk}
\end{align}
respectively. The rescaled Stefan condition is
$\d \check{s}/\d \check{t} = -\check{q}(\check{s}(\check{t}),\check{t})$.
The Newton boundary condition reduces to $\check{T}(0,\check{t}) = -1$.
Solving this system with the matching conditions
$\check{T} \sim -1 + \check{x}$ and $\check{s} \sim 1$ as $\check{t} \sim 0$
shows that the temperature is in its quasi-steady profile
given by $\check{T}(\check{x},\check{t}) = -1 + \check{x}/\check{s}(\check{t})$. 
Using this temperature profile
in the GK equation \eqref{large_bl_r5:gk} provides the solution for the flux,
$\check{q} = -1 / \check{s} + (1 / \check{s}^2)( \d \check{s} / \d \check{t} )$, which yields
a Stefan condition of the form
\begin{align}
  \td{\check{s}}{\check{t}} = \frac{\check{s}(\check{t})}{1 + \check{s}(\check{t})^2}.
\end{align}
Solving this equation with the initial condition $\check{s}(0) = 1$ 
results in an implicit expression for the solidification front,
\begin{align}
  \frac{1}{2}\check{s}(\check{t})^2 + \log \check{s}(\check{t}) = \check{t} + \frac{1}{2},
  \label{large_bl_r5:s}
\end{align}
which can be written in terms of the LambertW function 
as $\check{s}(\check{t}) = [\text{W}(\e^{2\check{t} + 1})]^{1/2}$. 

\begin{figure}
  \centering
  \subfigure[]{\includegraphics[width=0.49\textwidth]{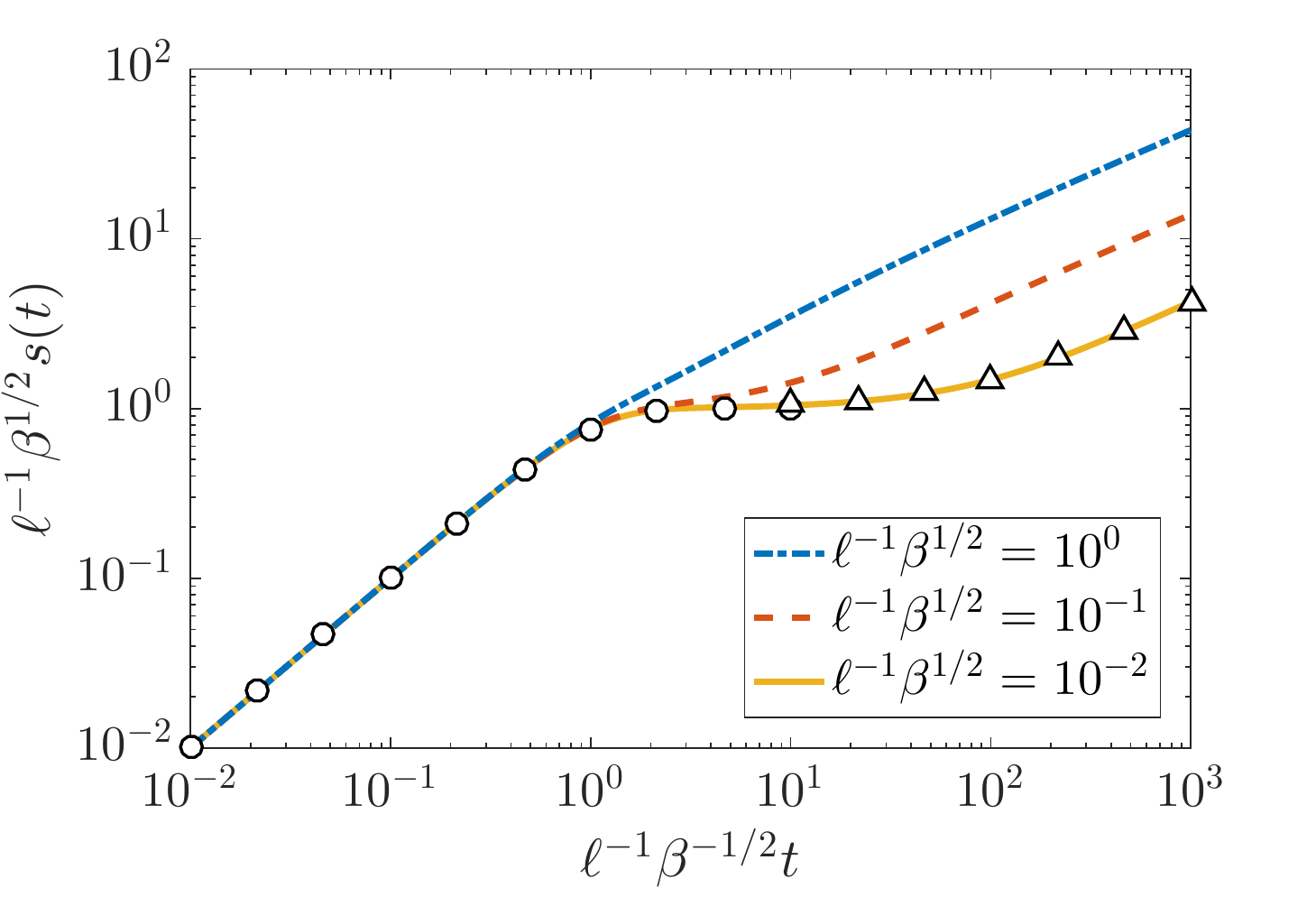}}
  \subfigure[]{\includegraphics[width=0.49\textwidth]{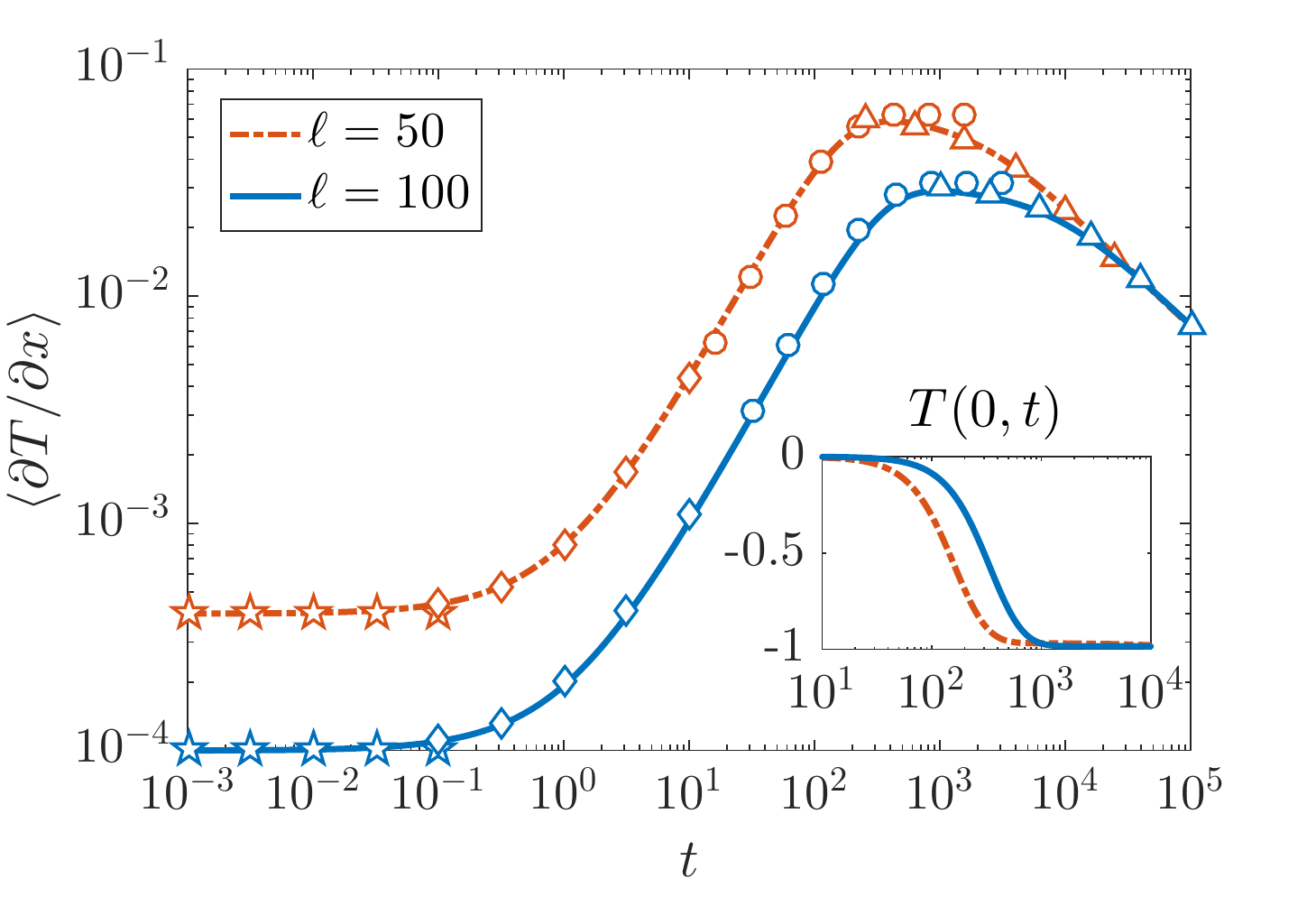}} \\
  \subfigure[]{\includegraphics[width=0.49\textwidth]{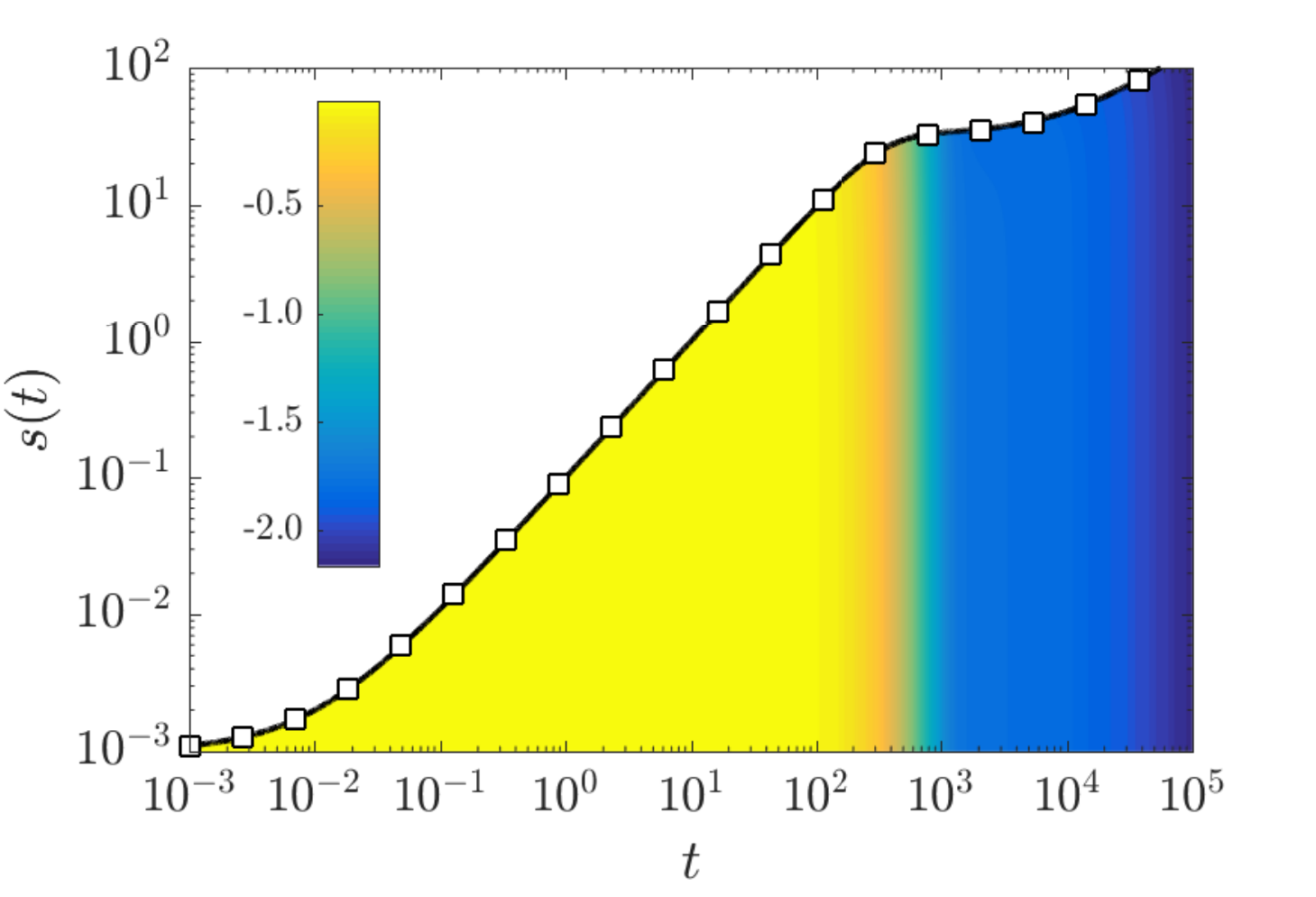}}
  \subfigure[]{\includegraphics[width=0.49\textwidth]{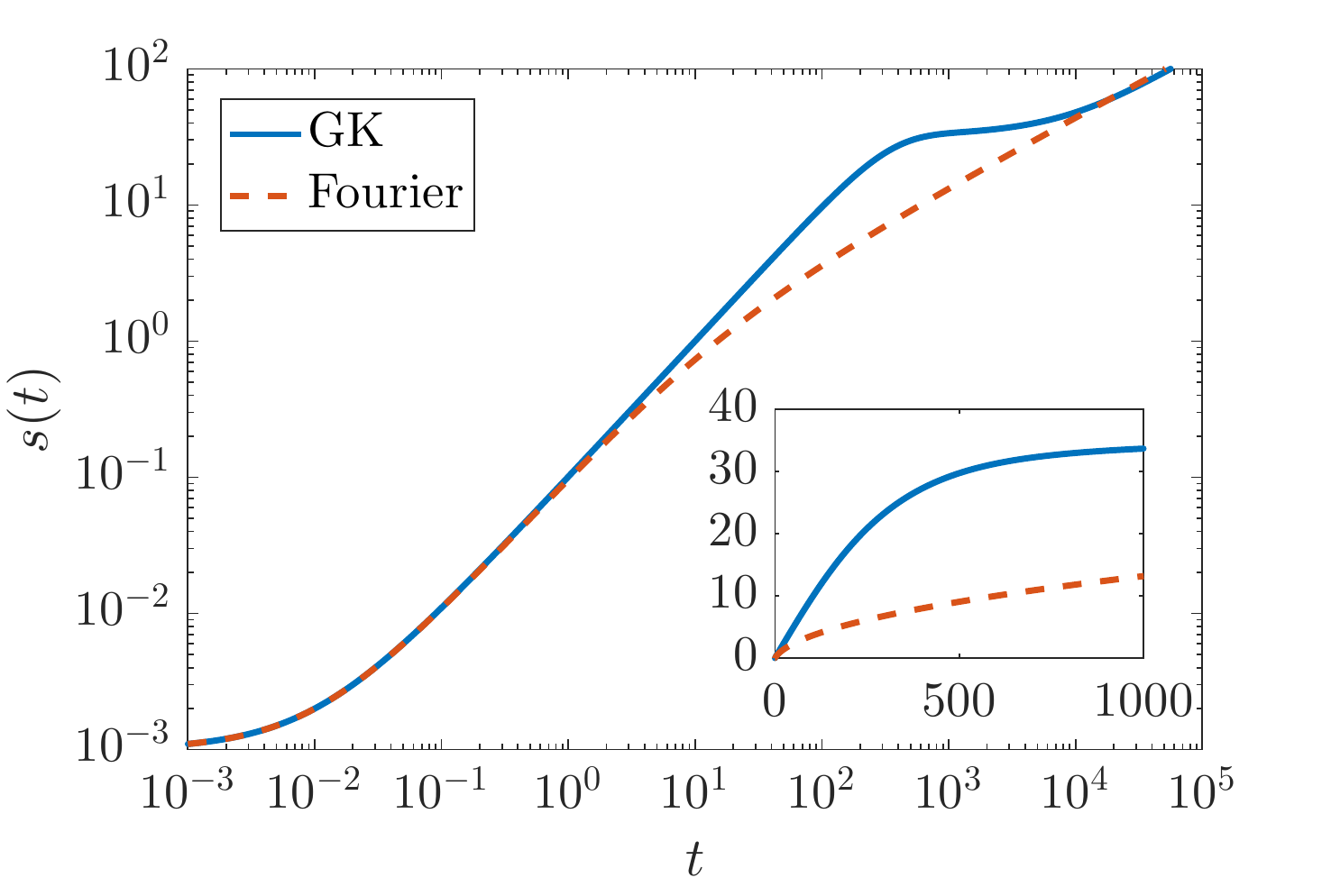}}
  \caption{Solidification dynamics in the limit of large MFP and Stefan number.
    (a): The blending of the fourth and fifth
    time regimes as the parameter $\ell^{-1} \beta^{1/2}$ increases.
    (b): Evolution of the mean temperature gradient (main panel) and the temperature 
      at the boundary in contact with the cold environment (inset).
    (c): Comparison of a numerical simulation of the full model (solid lines)
    with the composite asymptotic solution (squares) given by \eqref{large_bl:comp}.
    The superposed colour gradient represents the
    logarithm of the flux, $\log_{10} (-q)$. 
    (d): Evolution of the solidification front as predicted from the GK (solid)
    and Fourier (dashed) models. The inset and main panel are the same except 
    for the axes scalings.
    Lines represent numerical solutions of \eqref{eqn:Gs} while stars, diamonds, circles and
    triangles correspond to asymptotic solutions in the second, third, fourth, and fifth
    time regimes, respectively. 
    Parameter values used in panels (b) and (c) are $\gamma = 1$,
    $\varepsilon = 10^{-3}$, $\beta = 10$, and $\ell = 100$.}
  \label{fig:large_bl}
\end{figure}

Equations \eqref{large_bl_r4:s} and
\eqref{large_bl_r5:s} can be directly obtained
from \eqref{eqn:Gs} by considering a matched asymptotic
expansion in terms of $\alpha \ll 1$. Seeking
a solution of the form $G(t) = G_0(\bar{t}) + O(\alpha)$
and $\bar{s}(\bar{t}) = \bar{s}_0(\bar{t}) + O(\alpha)$ in \eqref{eqn:Gs} results in 
$G_0(t) = \bar{s}_0(\bar{t}) = \tanh \bar{t}$, i.e., \eqref{large_bl_r4:s}.
Rescaling time as $t = \alpha^{-1} \tau$ and writing
$G(t) = \tilde{G}_0(\tau) + O(\alpha)$ and $\bar{s}(t) = \tilde{s}_0(\tau) + O(\alpha)$
in \eqref{eqn:Gs} leads to the relations $\tilde{G}_0(\tau) = 1 / \tilde{s}_0(\tau)$ and
$\d (\tilde{G}_0 - \tilde{s}_0) / \d \tau = \tilde{G}_0$. These can be
combined into $\d \tilde{s}_0 / \d \tau = \tilde{s}_0 / (1 + \tilde{s}_0^2)$, which, upon
solving, yields \eqref{large_bl_r5:s}. In fact, this result is not surprising,
since the scales that have been used for time in the fourth and fifth
time regimes differ only by a factor of $\alpha^{-1}$. Therefore,
as $\alpha$ increases, the separation between the fourth and fifth time scales decreases. 
When $\alpha = O(1)$, these two regimes are blended together, as shown in Fig.~\ref{fig:large_bl} (a).

The evolution of the mean temperature gradient is shown in Fig.~\ref{fig:large_bl} (b) for two values of the
MFP $\ell$ with $\gamma = 1$, $\varepsilon = 10^{-3}$, and $\beta = 10$. Lines correspond to the numerical solutions
of the full model \eqref{eqn:full}; stars, diamonds, circles, and triangles represent the asymptotic solutions in
second, third, fourth, and fifth regimes, respectively. Due to the large effective thermal conductivity 
$\ell^2/\gamma$, the temperature gradient that is induced by the initial cooling of the solid in the first regime
(not shown) is very small.  The magnitude of the temperature gradient is consistent
with the scaling estimate of $\langle \pdf{T}{x}\rangle = O(\gamma\ell^{-2})$ obtained from \eqref{large_bl:12}.
As $t$ becomes $O(1)$ in size, the third time regime is entered and the temperature gradient undergoes a sustained
increase until the transition between the fourth ($t = O(l\beta^{1/2})$) and fifth ($t = O(\ell^2)$) time regimes.
The increase in temperature gradient can be conceptualised in terms of the effective thermal conductivity, which
decreases to the bulk value as the solid grows in size. The maximum gradient occurs when the region of solid in
contact with the cold environment obtains the same temperature as the cold environment, as shown in the inset of
Fig.~\ref{fig:large_bl} (b). The subsequent decrease in gradient in driven by the growth of the solid, since the
temperatures of the solid at its boundaries remain constant in time.

Figure \ref{fig:large_bl} (c) provides a comparison between the position of the solidification front
that has been computed from a numerical simulation of the full model \eqref{eqn:full} and the composite
asymptotic solution
\begin{align}
  s(t) = \varepsilon + \ell \beta^{-1/2} \tanh(\ell^{-1} \beta^{-1/2} t)
  + \ell \beta^{-1/2}[\text{W}(\e^{2t/\ell^2 + 1})]^{1/2}
  - \ell \beta^{-1/2}.
  \label{large_bl:comp}
\end{align}
The agreement between numerics and asymptotics is excellent. The figure
clearly shows a temporary decrease in the rate of front
propagation, which coincides with a marked decrease in the
magnitude of the thermal flux (shown as the superposed colour gradient).
This decrease in flux is due a transition between energy
transport mechanisms as the fifth time regime is entered.

The key difference between the fourth and fifth time regimes
is the dominant mechanism of energy transport. In the fourth time
regime, the flux is primarily driven by temporal changes
in the temperature gradient rather than the temperature gradient
itself; this is easily seen by
taking $\alpha \to 0$ in the leading-order part of
the GK equation \eqref{large_bl_r4:gk}. 
%As the temperature
%gradient settles into a slowly evolving quasi-steady state,
The decreasing growth of the temperature gradient
(see Fig.~\ref{fig:large_bl} (b)) diminishes the strength of the
primary mechanism of heat transport, leading to the
termination of the solidification process in this regime, which coincides
with the temperature gradient becoming stationary.
In the fifth time regime, the flux is driven by
the temperature gradient over large time scales;
see \eqref{large_bl_r5:gk}.
Thus, the temperature gradient provides a persistent mode of
energy transport that can sustain the solidification process,
even after the temperature profile settles into a slowly
evolving quasi-steady state.

Figure \ref{fig:large_bl} (d) compares the positions of the solidification
front that have been computed using the GK equation (solid line) and Fourier's law (dashed line)
for the thermal flux. The inset plots the same results using linear axes. 
Despite the slowing of front propagation that occurs in the GK model, strong
non-local effects lead to a prolonged period where the growth of the solid is approximately
linear, leading to a solidification process that is accelerated in
comparison to the case of pure Fourier conduction. Using the simulation 
parameters $\ell = 100$ and $\beta = 10$, the transition from
linear growth kinetics occurs in the GK model when $t \simeq 316 = O(l \beta^{1/2})$;
in the case of Fourier's law, this occurs when $t \simeq 10 = O(\beta)$. 
Interestingly, the gains that are made in the solidification process during the extended period
of linear growth are almost exactly offset by the slowing that occurs during the transition
into the diffusion-dominated third time regime. From Fig.~\ref{fig:large_bl} (d), it is seen
that the two curves eventually collapse onto each other, indicating the recovery of the usual
quasi-steady $s \sim (2t)^{1/2}$ solidification kinetics.

%================================================

\subsection{Limit of large relaxation time}
\label{sec:large_bg}

We now study the dynamics in the asymptotic limit $\gamma \to \infty$ and $\beta \to \infty$,
focusing on the case
when $\beta \ll \gamma$ with $\varepsilon \ll \gamma^{-1}$, which leads to substantial departures from
classical solidification kinetics. 

The problem can be decomposed into five distinct time regimes. The first two, defined by
$t = O(\gamma \varepsilon^2)$ and $t = O(\beta \varepsilon)$, describe the dynamics that occur
on length scales associated with the seed crystal, namely, the initial transport of heat to
the environment and growth of the solid. The third time regime occurs when $t = O(\beta \gamma^{-1})$
and, although the dynamics are dominated by non-classical effects, the thermal flux is driven by
temperature gradients. This results in a change of solidification kinetics from
$s \sim t$ to $s \sim t^{1/2}$. In the fourth time regime, $t = O(1)$, the flux becomes dependent on the
temperature gradient as well as the history of the temperature gradient, resulting in further
changes to the solidification kinetics. Finally, in the fifth regime, $t = O(\gamma)$, non-local
effects become negligible and the model reduces to a quasi-static version of the hyperbolic heat equation. 
On even longer time scales, $t \gg \gamma$, the system tends to the classical limit.

\emph{First and second time regimes:} The first two time regimes
are mathematically identical to those previously considered when $\gamma = O(1)$. The limiting
behaviour of the solutions is given by
$T \sim \gamma \ell^{-2}(x - s(t))$, $q \sim -1$, and $s \sim \beta^{-1} t$ for
$\beta \varepsilon \ll t \ll \beta \gamma^{-1}$. The upper bound on time ensures that the 
temperature is small in these two regimes. 

\emph{Third time regime}: The third time regime, $t = O(\beta \gamma^{-1})$, accounts for $O(1)$ 
temperatures and can be identified from the limiting behaviour in the second time
regime. Balancing terms in the Newton condition shows that $q = O(1)$, which, in turn, can be
used in the Stefan condition to obtain a length scale of $O(\gamma^{-1})$. Thus, the variables are
rescaled according to
$t = \beta \gamma^{-1} \bar{t}$, $x = \gamma^{-1} \bar{x}$, $s = \gamma^{-1} \bar{s}$,
$T = \bar{T}$, and $q = \bar{q}$, resulting in bulk equations given by
\subeq{
\begin{align}
  \pd{\bar{q}}{\bar{x}} &= O(\beta^{-1}), \\
  \pd{\bar{q}}{\bar{t}} + \beta \gamma^{-1} \pd{\bar{T}}{\bar{x}} &= -\ell^2 \pd{}{\bar{t}}\left(\pd{\bar{T}}{\bar{x}}\right) + O(\beta \gamma^{-2}).
  \label{bar:lgb_r3_GK}
\end{align}
}
The boundary and Stefan conditions are
$\bar{q}(0,\bar{t}) = -(1 + \bar{T}(0,\bar{t}))$,
$\bar{T}(\bar{s}(\bar{t}),\bar{t}) = 0$, and
$\d \bar{s} / \d \bar{t} = -\bar{q}(\bar{s}(\bar{t}),\bar{t})$,
with matching conditions given by
$\bar{T} \sim \ell^{-2}(\bar{x} - \bar{s}(\bar{t}))$ and
$\bar{s} \sim \bar{t}$ as $\bar{t} \sim 0$. The leading-order part of the GK equation
\eqref{bar:lgb_r3_GK} can be integrated in time, yielding $\bar{q} = -\ell^2 \pdf{\bar{T}}{\bar{x}}$,
thus recovering a form of Fourier's law. 
Upon solving a classical problem for the temperature, one finds that
\subeq{
\label{bar:lgb_r3_Tq}
\begin{align}
  \bar{T}(\bar{x},\bar{t}) = \frac{\bar{x} - \bar{s}(\bar{t})}{\bar{s}(\bar{t})^2 + \ell^2},
  \quad
  \bar{q}(\bar{x},\bar{t}) = -\frac{\ell^2}{\bar{s}(\bar{t})^2 + \ell^2},
\end{align}
where the position of the solidification front is determined from the positive root of
\begin{align}
  2 \bar{s}(\bar{t}) + \ell^{-2} \bar{s}(\bar{t})^2 = 2 \bar{t}. \label{bar:lbg_r3_s}
\end{align}
}
The small- and large-time limits of \eqref{bar:lbg_r3_s} are $\bar{s} \sim \bar{t}$ and
$\bar{s} \sim \ell (2 \bar{t})^{1/2}$, respectively, which indicate that a reduction in the rate
of solidification occurs in this time regime. This slowing is attributed to the weakening of the
temperature gradient and thus the thermal flux, as seen from the limits
$\bar{T} \sim -1 + \bar{x} / \bar{s}(\bar{t})$ and 
$\bar{q} \sim -\ell^2 / \bar{s}(\bar{t})$ for $\bar{t} \gg 1$. 

\emph{Fourth time regime}: The diminishing flux and temperature gradient result in additional
transport mechanisms becoming relevant in the GK equation \eqref{bar:lgb_r3_GK}. The limiting behaviour
of the solution in the third time regime determines the scales for space, temperature, and flux.
The time scale of $t = O(1)$ is chosen so that the temperature gradient appears
in the leading-order GK equation. The corresponding rescaling is given by
$t = \tilde{t}$, 
$x = \beta^{-1/2} \gamma^{-1/2} \tilde{x}$,
$s = \beta^{-1/2} \gamma^{-1/2} \tilde{s}$,
$T = \tilde{T}$, and $q = \beta^{1/2}\gamma^{-1/2}\tilde{q}$.
The rescaled energy balance and GK equation are given by
\subeq{
\begin{align}
  \pd{\tilde{q}}{\tilde{x}} &= O(\beta^{-1}), \label{lbg:r2_energy} \\
  \pd{\tilde{q}}{\tilde{t}} + \gamma^{-1}\tilde{q} + \pd{\tilde{T}}{\tilde{x}} &= 
  -\ell^2 \pd{}{\tilde{t}}\left(\pd{\tilde{T}}{\tilde{x}}\right), \label{lbg:r2_gk}
\end{align}
}
which have boundary conditions $\tilde{T}(0,\tilde{t}) = -1 + O(\beta^{1/2}\gamma^{-1/2})$ and
$\tilde{T}(\tilde{s}(\tilde{t}),\tilde{t}) = 0$. The
Stefan condition is $\d \tilde{s} / \d \tilde{t} = -\tilde{q}(s(\tilde{t}),\tilde{t})$. 
Matching conditions are
$\tilde{T} \sim -1 + \tilde{x} / \tilde{s}(\tilde{t})$,
$\tilde{q} \sim -\ell (2 \tilde{t})^{-1/2}$, and
$\tilde{s} \sim \ell (2 \tilde{t})^{1/2}$ as $\tilde{t} \sim 0$. Integrating the leading-order
GK equation \eqref{lbg:r2_gk} in time results in
\begin{align}
  \tilde{q} = -\ell^2 \pd{\tilde{T}}{\tilde{x}} - \int_{0}^{\tilde{t}} \pd{\tilde{T}}{\tilde{x}}\,\d
\tilde{t}',
\end{align}
clearly showing the dependence of the flux on the history of the temperature gradient. 
To make further progress, the GK equation \eqref{lbg:r2_gk} is differentiated with respect to
$\tilde{x}$ and the energy equation \eqref{lbg:r2_energy} together with the boundary and
matching conditions are used to show that the temperature is
$\tilde{T}(\tilde{x},\tilde{t}) = -1 + \tilde{x}/\tilde{s}(\tilde{t})$.  
Inserting the expression for the temperature in the leading-order part of the GK equation
\eqref{lbg:r2_gk} leads to a coupled system of equations for the flux and position of the
solidification front:
\subeq{
  \label{lbg:r4_red}
  \begin{align}
    \td{\tilde{q}}{\tilde{t}} &= -\frac{1}{\tilde{s}} - \left(\frac{\ell}{\tilde{s}}\right)^2\tilde{q}, 
                                \label{lbg:r4_red_q}
    \\
    \td{\tilde{s}}{\tilde{t}} &= -\tilde{q},
  \end{align}
}
with $\tilde{q} \sim -\ell (2 \tilde{t})^{-1/2}$ and
$s \sim \ell (2\tilde{t})^{1/2}$ for $\tilde{t} \sim 0$. For small times, \eqref{lbg:r4_red_q} reduces
to $\d \tilde{q}/\d \tilde{t} \sim -(l/\tilde{s})^2 \tilde{q}$, showing that the flux increases in time
(recall that $\tilde{q} < 0$). However, for large times, $\d \tilde{q}/\d \tilde{t} \sim -1/\tilde{s}$, 
corresponding to a decreasing flux. The non-monotonicity of the flux corresponds to a change in the
dominant mechanism of thermal transport. 
For $\tilde{t} \ll 1$, the flux is driven by the temperature
gradient; for $\tilde{t} \gg 1$, it is driven by memory effects and the history of the 
temperature gradient, leading to markedly different long-term solidification
kinetics given by $\d \tilde{s} / \d \tilde{t} \sim [2 \log(\tilde{s} / S)]^{1/2}$, where $S = \ell^2 (\pi / 2)^{1/2}$.

\emph{Fifth time regime}: The sustained growth of the solid beyond the dimensions of the phonon
MFP reduces the influence of non-local effects.  The fifth time regime describes the evolution of the
solidification process in the absence of such effects.  
The associated scales for the variables can be determined from the GK equation in
the fourth time regime \eqref{lbg:r2_gk}, which shows that a different balance occurs when 
$\tilde{x}, \tilde{t} = O(\gamma)$. The original dimensionless variables are therefore scaled as
$t = \gamma \hat{t}$, $x = \beta^{-1/2}\gamma^{1/2} \hat{x}$, $s = \beta^{-1/2}\gamma^{1/2} \hat{s}$,
$T = \hat{T}$, and $q = \beta^{1/2}\gamma^{-1/2} \hat{q}$, leading to bulk equations given by
\subeq{
  \label{lbg:r5}
  \begin{align}
  \pd{\hat{q}}{\hat{x}} &= O(\beta^{-1}), \label{lbg:r5_energy} \\
  \pd{\hat{q}}{\hat{t}} + \hat{q} + \pd{\hat{T}}{\hat{x}} &= 
  O(\gamma^{-1}), \label{lbg:r5_gk}
  \end{align}
}
corresponding to the quasi-static hyperbolic heat equation. The leading-order boundary conditions are
$\hat{T}(0,\hat{t}) = -1$ and 
$\hat{T}(\hat{s}(\hat{t}),\hat{t}) = 0$. Differentiation of \eqref{lbg:r5_gk}
with respect to $\hat{x}$ and using \eqref{lbg:r5_energy} along with the boundary conditions shows that
$\hat{T} = -1 + \hat{x} / \hat{s}(\hat{t})$. The problem then reduces to a system of differential
equations for the flux and position of the solidification front given by
\subeq{
  \label{lbg:r5_red}
  \begin{align}
    \td{\hat{q}}{\hat{t}} + \hat{q} &= -\frac{1}{\hat{s}}, \\
    \td{\hat{s}}{\hat{t}} &= -\hat{q}.
  \end{align}
}
For $\hat{t} \ll 1$, we find that $\d \hat{q} / \d \hat{t} \sim -1 / \hat{s}$, 
consistent with the limiting
behaviour in the fourth regime. For large times, $\hat{t} \gg 1$, we have $\hat{q} \sim -1/\hat{s}$
and thus $\hat{s} \sim (2 \hat{t})^{1/2}$, recovering the classical solidification kinetics as the
flux once again becomes primarily driven by temperature gradients.

\begin{figure}
  \centering
  \subfigure[]{\includegraphics[width=0.49\textwidth]{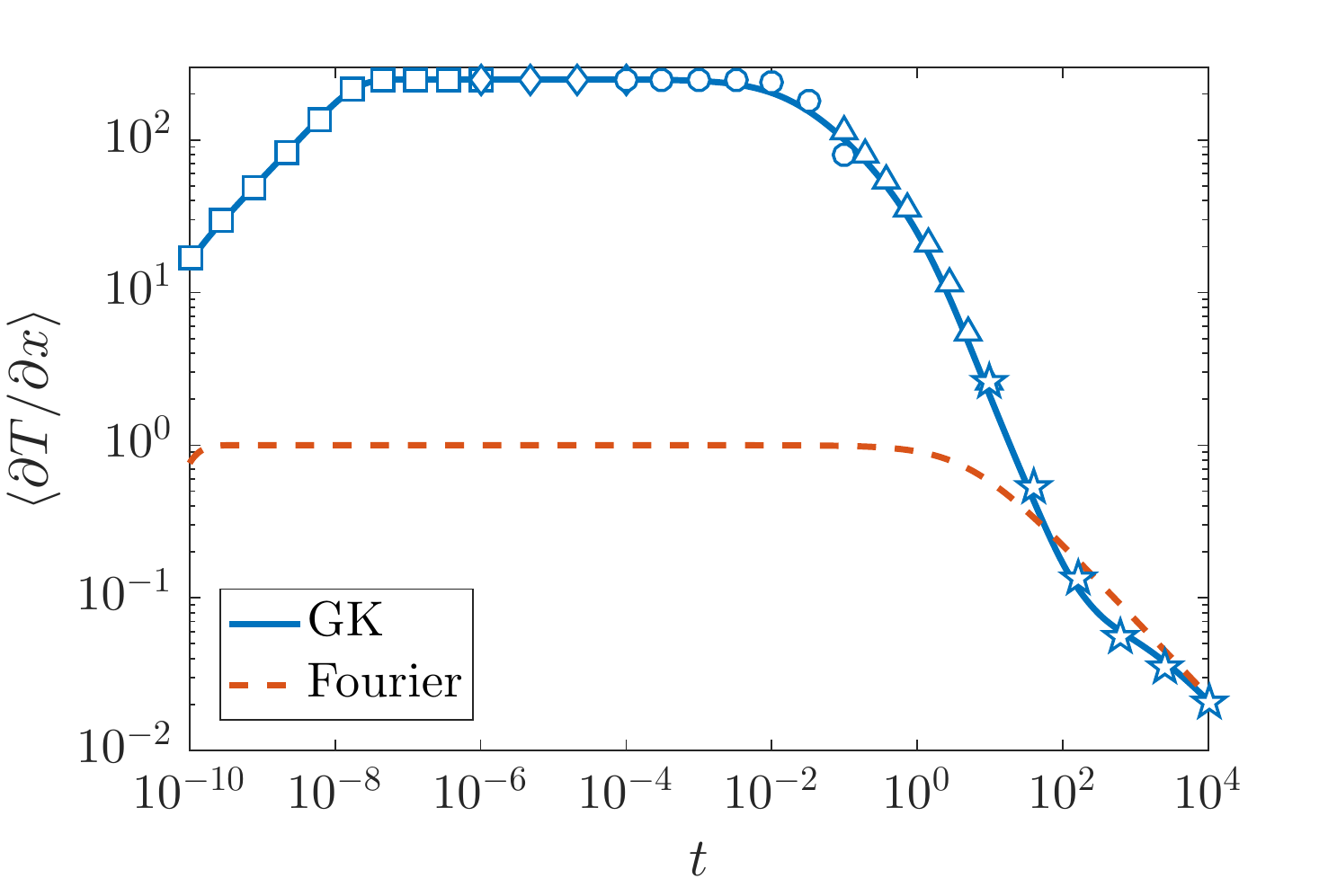}}
  \subfigure[]{\includegraphics[width=0.49\textwidth]{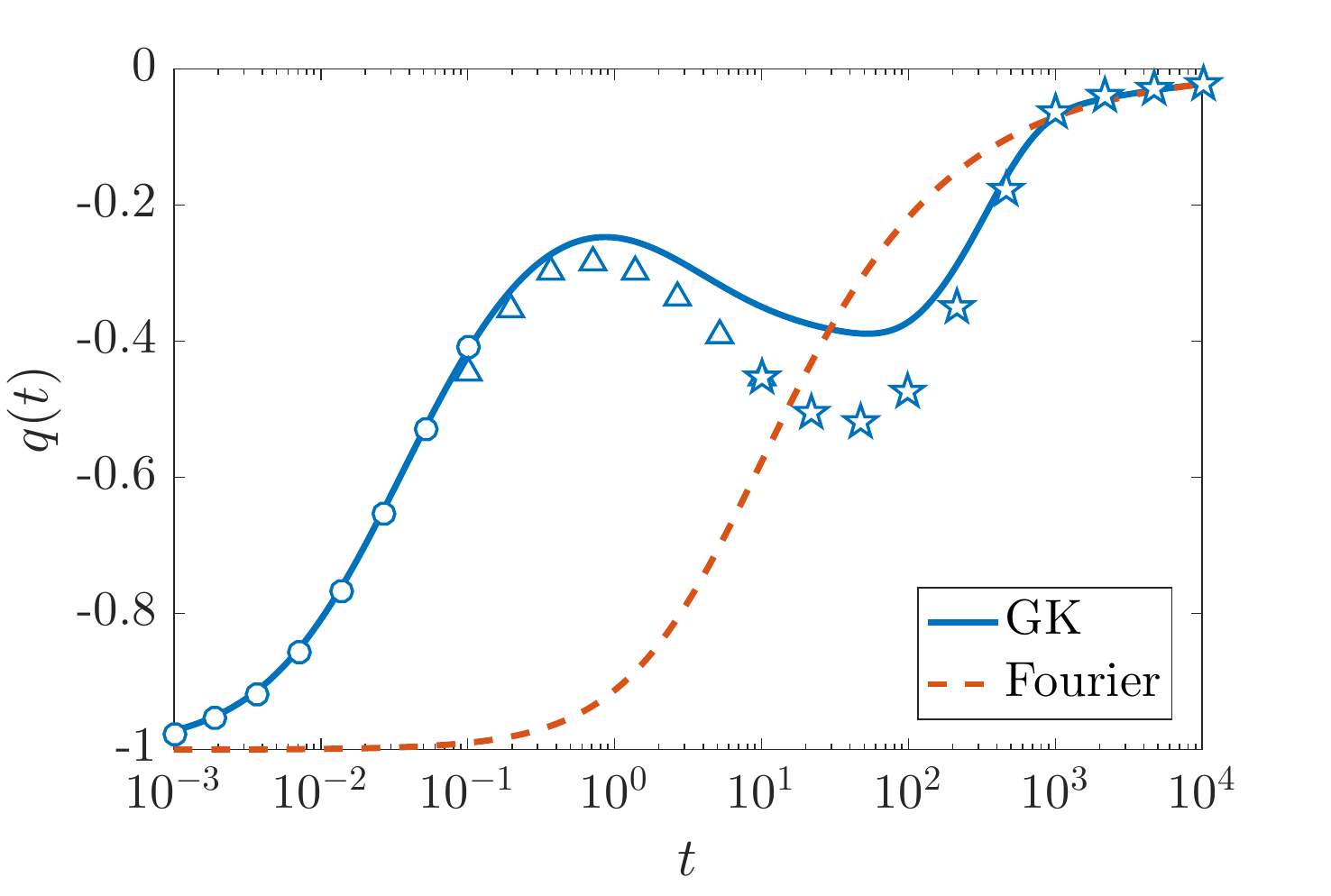}}
  \subfigure[]{\includegraphics[width=0.49\textwidth]{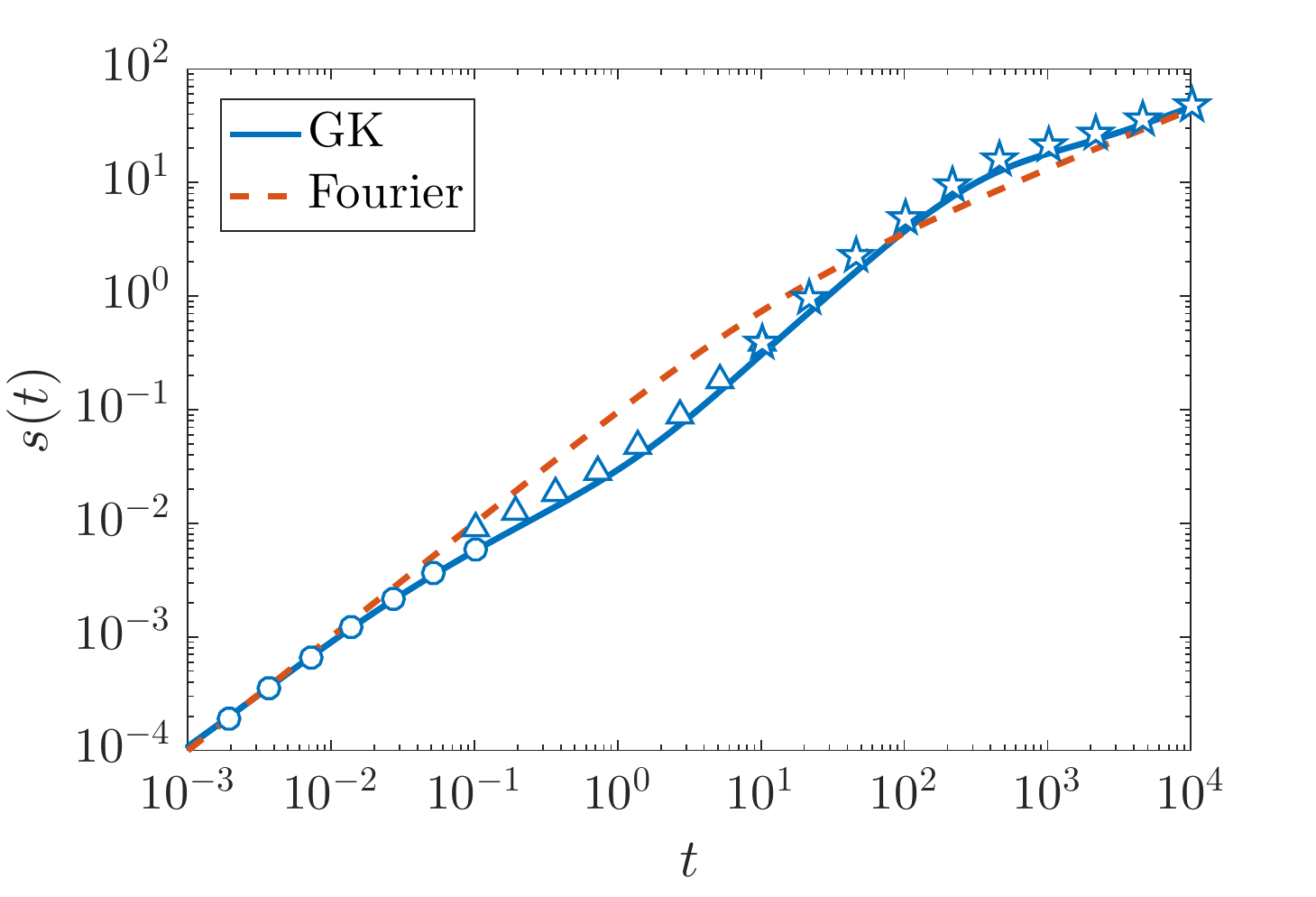}}
  \caption{Solidification dynamics in the limit of large relaxation time and
    Stefan number. Comparisons of the temperature gradient (a), flux (b),
    and position of the solidification front
    (c) from the GK (solid line) and Fourier (dashed line) models. 
    Diamonds and circles denote asymptotic solutions in the second and third regimes;
    squares, triangles, and stars represent numerical solutions of the 
    reduced models in the first, fourth, and fifth regimes given by \eqref{pm:r0_thermal}, 
    \eqref{lbg:r4_red}, and \eqref{lbg:r5_red}, respectively.
    Note that $q(t)$ is calculated from the full model by taking the integrated
    average of $q(x,t)$ in space.
    Parameter values are $\beta = 10$, $\gamma = 250$, $\ell = 1$, and $\varepsilon = 10^{-5}$.}
  \label{fig:large_beta_gamma}
\end{figure}

Figure \ref{fig:large_beta_gamma} shows the evolution of the temperature gradient (panel(a)),
flux (panel (b)), and the position of the solidification front (panel (c))
in the case of $\beta = 10$, $\gamma = 250$, $\ell = 1$, and $\varepsilon = 10^{-5}$. Although
these parameters lie outside of the estimates given in Sec.~\ref{sec:params}, they provide sufficient
time-scale separation to illustrate the range of behaviour that occurs across the various 
regimes. In Fig.~\ref{fig:large_beta_gamma}, the solid line corresponds to the
numerical solution of the full model \eqref{eqn:full}.  Diamonds and circles denote 
the asymptotic solutions in the second and third regime, and squares, triangles and stars represent
numerical solutions of the reduced model in the first, fourth, and fifth regime, given by
\eqref{pm:r0_thermal}, \eqref{lbg:r4_red} and \eqref{lbg:r5_red}, respectively. Obtaining a precise
matching condition for \eqref{lbg:r5_red} was not possible and thus it was patched to the numerical
solution in the fourth regime. The asymptotic solutions capture the qualitative trend
seen in the numerical simulation and the error is consistent with the magnitude of the
$O(\beta^{1/2}\gamma^{-1/2})$ terms that are neglected when analysing the model in
the fourth time regime. Despite these errors, the asymptotic solution 
for the position of the solidification front is in good agreement with the numerical simulation.

The dashed lines in Fig.~\ref{fig:large_beta_gamma} correspond to the temperature gradient, flux, and 
position of the solidification front computed from Fourier's law and highlight the vastly different
dynamics that occur when the relaxation time is large. 
The small effective thermal conductivity $\ell^2/\gamma$ leads to the rapid development of large temperature
gradients in the first time regime, $t \simeq 10^{-8} = O(\gamma\varepsilon^2)$, which are constant throughout
the second time regime, $t \simeq 10^{-4} = O(\beta\varepsilon)$.
The third time regime, $t \simeq 10^{-2} = O(\beta \gamma^{-1})$, 
captures the decrease in thermal flux that occurs as a result of the
poor effective thermal conductivity.  As Fig.~\ref{fig:large_beta_gamma} (c) shows, the diminished
transport of thermal energy in the third time regime leads to an 
earlier transition to slower $s \sim t^{1/2}$ solidification kinetics compared to the Fourier case. 
However, as the system enters the fourth time regime, $t = O(1)$, the history of the temperature gradient becomes 
similar in magnitude to the instantaneous value of the temperature gradient, and this extra
contribution to the thermal flux leads to enhanced energy transport and a sustained
acceleration of the solidification process. In the fifth time regime, $t \simeq 10^{2} = O(\gamma)$,
these memory effects become
sub-dominant and the flux, once again, becomes driven by temperature gradients with a thermal
conductivity that coincides with the bulk value, recovering the classical solidification kinetics.

\section{Discussion and conclusion}
\label{sec:conc}

The asymptotic analysis reveals that a form of Fourier's law, with an effective thermal conductivity depending 
on the non-classical parameters, is recovered on small time scales. 
On large time scales, the weakening of non-classical transport
mechanisms leads to a recovery of the classical form of Fourier's law, based on the bulk thermal conductivity of 
the material. These results highlight the size-dependent nature of the thermal conductivity of nanomaterials and
support the idea that, in certain regimes, nanoscale heat conduction can be described by
Fourier's law if a suitable choice of thermal conductivity is made \cite{alvarez2009, dong2014, ma2012, calvo2018, sobolev2014}.

Key differences in the dominant transport mechanism across intermediate time scales are observed when
the relaxation time and phonon MFP are large. 
In the former case, the flux is driven by the history (integral in time) of the
temperature gradient. This is in perfect contrast to the latter case, where the flux is driven by the
instantaneous rate of change (time derivative) of the temperature gradient. Despite these seemingly
opposite relationships between the flux and temperature gradient, both lead to an enhancement of 
energy transport across the solid and a non-trivial increase in the rate of solidification. 

% Overall, our results indicate that non-Fourier heat conduction can lead to marked differences 
% in the solidification dynamics if the dimensionless relaxation time or phonon MFP are large. 
% These conditions ensure that non-classical transport mechanisms are still active by the time
% solidification begins and are therefore linked to the assumption of a large Stefan number,
% which is characteristic of slow phase-change processes. 
% In rapidly solidifying systems with Stefan numbers that are $O(1)$ or smaller, 
% non-Fourier heat conduction is 
% likely to play an increased role due to the tighter coupling between the solidification and transport
% processes. \note{Add discussion of $\beta = O(1)$ stuff}

The focus on solidification in the limit of large Stefan number is to facilitate the asymptotic analysis. 
However, a numerical study of the model with $O(1)$ Stefan numbers indicates the same qualitative
behaviours occur. This similarity is explained by the fact that $O(1)$ Stefan numbers
do not change the ordering of the time regimes identified by the asymptotic analysis. 
Physically, it means that the rate of solidification does not alter the
sequence of dominant thermal transport mechanisms; rather, it only affects when the transition between these
mechanisms takes place.

In classical models of phase change, the governing equations for melting and solidification are the same. Thus,
theoretical insights into one process can be applied to the other. The GK equation, which can be derived from the
framework of extended irreversible thermodynamics \cite{jou2010}, in principle, can be used to describe the flux of
thermal energy in the liquid phase. However, much less is known about the precise mechanism of thermal transport in
nanoscale liquids and phonon-based theories are still being developed \cite{bolmatov2012}. A difference between
thermal transport mechanisms in solids and liquids would break the analogy between solidification and melting,
thereby preventing the results from this study to be applied to the problem of nanoscale melting.

By studying a simple model with asymptotic techniques, we are able to obtain novel insights into how
solidification is affected by non-Fourier heat conduction. Although motivated by applications in 
nanotechnology, the model and analysis can be used more generally to describe rapid solidification processes
occurring on time and length scales that are commensurate with the thermal relaxation time and phonon mean
free path of a material.

\section*{Acknowledgements}

We thank Brian Wetton for his advice regarding the numerical solution of this problem. 
This project has received funding from the European Union's Horizon 2020 research and innovation
programme under grant agreement No 707658. MC acknowledges that the research leading to these results
has received funding from `la Caixa' Foundation.
TGM acknowledges the support of a Ministerio de Ciencia e Innovaci\'on grant MTM2014-56218.
The authors have been partially funded by the CERCA Programme of the Generalitat de Catalunya.

\bibliographystyle{unsrtnat}
\bibliography{refs}

\begin{thebibliography}{65}
\providecommand{\natexlab}[1]{#1}
\providecommand{\url}[1]{\texttt{#1}}
\expandafter\ifx\csname urlstyle\endcsname\relax
  \providecommand{\doi}[1]{doi: #1}\else
  \providecommand{\doi}{doi: \begingroup \urlstyle{rm}\Url}\fi

\bibitem[Garnett et~al.(2011)Garnett, Brongersma, Cui, and
  McGehee]{garnett2011}
E.~C. Garnett, M.~L. Brongersma, Y.~Cui, and M.~D. McGehee.
\newblock Nanowire solar cells.
\newblock \emph{Annu.~Rev.~Mater.~Res.}, 41:\penalty0 269--295, 2011.

\bibitem[Wang(2005)]{wang2005}
J.~Wang.
\newblock Carbon-nanotube based electrochemical biosensors: {A} review.
\newblock \emph{Electroanal.}, 17\penalty0 (1):\penalty0 7--14, 2005.

\bibitem[Salata(2004)]{salata2004}
O.~V. Salata.
\newblock Applications of nanoparticles in biology and medicine.
\newblock \emph{J.~Nanobiotechnol}, 2\penalty0 (1):\penalty0 3, 2004.

\bibitem[Mah et~al.(2000)Mah, Zolotukhin, Fraites, Dobson, Batich, and
  Byrne]{mah2000}
C.~Mah, I.~Zolotukhin, T.~J. Fraites, J.~Dobson, C.~Batich, and B.~J. Byrne.
\newblock Microsphere-mediated delivery of recombinant {AAV} vectors in vitro
  and in vivo.
\newblock \emph{Mol.~Ther.}, 1\penalty0 (5):\penalty0 S239--S242, 2000.

\bibitem[Nam et~al.(2003)Nam, Thaxton, and Mirkin]{nam2003}
J.-M. Nam, C.~S. Thaxton, and C.~A. Mirkin.
\newblock Nanoparticle-based bio-bar codes for the ultrasensitive detection of
  proteins.
\newblock \emph{Science}, 301\penalty0 (5641):\penalty0 1884--1886, 2003.

\bibitem[Ma et~al.(2003)Ma, Wong, Kong, and Peng]{ma2003}
J.~Ma, H.~Wong, L.~B. Kong, and K.-W. Peng.
\newblock Biomimetic processing of nanocrystallite bioactive apatite coating on
  titanium.
\newblock \emph{Nanotechnology}, 14\penalty0 (6):\penalty0 619, 2003.

\bibitem[Pop et~al.(2006)Pop, Sinha, and Goodson]{pop2006}
E.~Pop, S.~Sinha, and K.~E. Goodson.
\newblock Heat generation and transport in nanometer-scale transistors.
\newblock \emph{P.~IEEE}, 94\penalty0 (8):\penalty0 1587--1601, 2006.

\bibitem[Poudel et~al.(2008)Poudel, Hao, Ma, Lan, Minnich, Yu, Yan, Wang, Muto,
  Vashaee, Chen, Liu, Dresselhaus, Chen, and Ren]{poudel2008}
B.~Poudel, Q.~Hao, Y.~Ma, Y.~Lan, A.~Minnich, B.~Yu, X.~Yan, D.~Wang, A.~Muto,
  D.~Vashaee, X.~Chen, J.~Liu, M.~S. Dresselhaus, G.~Chen, and Z.~Ren.
\newblock High-thermoelectric performance of nanostructured bismuth antimony
  telluride bulk alloys.
\newblock \emph{Science}, 320\penalty0 (5876):\penalty0 634--638, 2008.
\newblock ISSN 0036-8075.
\newblock \doi{10.1126/science.1156446}.
\newblock URL \url{http://science.sciencemag.org/content/320/5876/634}.

\bibitem[Hamad-Schifferli et~al.(2002)Hamad-Schifferli, Schwartz, Santos,
  Zhang, and Jacobson]{hamad2002}
K.~Hamad-Schifferli, J.~J. Schwartz, A.~T. Santos, S.~Zhang, and J.~M.
  Jacobson.
\newblock Remote electronic control of dna hybridization through inductive
  coupling to an attached metal nanocrystal antenna.
\newblock \emph{Nature}, 415\penalty0 (6868):\penalty0 152--155, 2002.

\bibitem[Cahill et~al.(2003)Cahill, Ford, Goodson, Mahan, Majumdar, Maris,
  Merlin, and Phillpot]{cahill2003}
D.~G. Cahill, W.~K. Ford, K.~E. Goodson, G.~D. Mahan, A.~Majumdar, H.~J. Maris,
  R.~Merlin, and S.~R. Phillpot.
\newblock Nanoscale thermal transport.
\newblock \emph{J.~Appl.~Phys.}, 93\penalty0 (2):\penalty0 793--818, 2003.

\bibitem[Siemens et~al.(2010)Siemens, Li, Yang, Nelson, Anderson, Murnane, and
  Kapteyn]{siemens2010}
M.~E. Siemens, Q.~Li, R.~Yang, K.~A. Nelson, E.~H. Anderson, M.~M. Murnane, and
  H.~C. Kapteyn.
\newblock Quasi-ballistic thermal transport from nanoscale interfaces observed
  using ultrafast coherent soft x-ray beams.
\newblock \emph{Nat.~Mater.}, 9\penalty0 (1):\penalty0 26--30, 2010.

\bibitem[Cahill et~al.(2014)Cahill, Braun, Chen, Clarke, Fan, Goodson,
  Keblinski, King, Mahan, Majumdar, Maris, Phillpot, Pop, and Shi]{cahill2014}
D.~G. Cahill, P.~V. Braun, G.~Chen, D.~R. Clarke, S.~Fan, K.~E. Goodson,
  P.~Keblinski, W.~P. King, G.~D. Mahan, A.~Majumdar, H.~J. Maris, S.~R.
  Phillpot, E.~Pop, and L.~Shi.
\newblock Nanoscale thermal transport. ii. 2003--2012.
\newblock \emph{Appl.~Phys.~Rev.}, 1\penalty0 (1):\penalty0 011305, 2014.

\bibitem[Nie et~al.(2011)Nie, Liu, Dong, and Wang]{nie2011}
A.~Nie, J.~Liu, C.~Dong, and H.~Wang.
\newblock Electrical failure behaviors of semiconductor oxide nanowires.
\newblock \emph{Nanotechnology}, 22\penalty0 (40):\penalty0 405703, 2011.

\bibitem[Buffat and Borel(1976)]{buffat1976}
P.~Buffat and J.~P. Borel.
\newblock Size effect on the melting temperature of gold particles.
\newblock \emph{Phys.~Rev. A}, 13\penalty0 (6):\penalty0 2287, 1976.

\bibitem[David et~al.(1995)David, Lereah, Deutscher, Kofman, and
  Cheyssac]{david1995}
T.~B. David, Y.~Lereah, G.~Deutscher, R.~Kofman, and P.~Cheyssac.
\newblock Solid-liquid transition in ultra-fine lead particles.
\newblock \emph{Philos.~Mag.~A}, 71\penalty0 (5):\penalty0 1135--1143, 1995.

\bibitem[Wronski(1967)]{wronski1967}
C.~R.~M. Wronski.
\newblock The size dependence of the melting point of small particles of tin.
\newblock \emph{Brit.~J.~Appl.~Phys.}, 18\penalty0 (12):\penalty0 1731, 1967.

\bibitem[Lai et~al.(1996)Lai, Guo, Petrova, Ramanath, and Allen]{lai1996}
S.~L. Lai, J.~Y. Guo, V.~Petrova, G.~Ramanath, and L.~H. Allen.
\newblock Size-dependent melting properties of small tin particles:
  nanocalorimetric measurements.
\newblock \emph{Phys.~Rev.~Lett}, 77\penalty0 (1):\penalty0 99, 1996.

\bibitem[Zhang et~al.(2000)Zhang, Efremov, Schiettekatte, Olson, Kwan, Lai,
  Wisleder, Greene, and Allen]{zhang2000}
M.~Zhang, M.~Y. Efremov, F.~Schiettekatte, E.~A. Olson, A.~T. Kwan, S.~L. Lai,
  T.~Wisleder, J.~E. Greene, and L.~H. Allen.
\newblock Size-dependent melting point depression of nanostructures:
  nanocalorimetric measurements.
\newblock \emph{Phys.~Rev.~B}, 62\penalty0 (15):\penalty0 10548, 2000.

\bibitem[Sun and Simon(2007)]{sun2007}
J.~Sun and S.~L. Simon.
\newblock The melting behavior of aluminum nanoparticles.
\newblock \emph{Thermochim.~Acta}, 463\penalty0 (1):\penalty0 32--40, 2007.

\bibitem[Tolman(1949)]{tolman1949}
R.~C. Tolman.
\newblock The effect of droplet size on surface tension.
\newblock \emph{J.~Chem.~Phys.}, 17\penalty0 (3):\penalty0 333--337, 1949.

\bibitem[Chang et~al.(2008)Chang, Okawa, Garcia, Majumdar, and
  Zettl]{chang2008}
C.-W. Chang, D.~Okawa, H.~Garcia, A.~Majumdar, and A.~Zettl.
\newblock Breakdown of {F}ourier's law in nanotube thermal conductors.
\newblock \emph{Phys.~Rev.~Lett.}, 101\penalty0 (7):\penalty0 075903, 2008.

\bibitem[Li et~al.(2003)Li, Wu, Kim, Shi, Yang, and Majumdar]{li2003}
D.~Li, Y.~Wu, P.~Kim, L.~Shi, P.~Yang, and A.~Majumdar.
\newblock Thermal conductivity of individual silicon nanowires.
\newblock \emph{Appl.~Phys.~Lett.}, 83\penalty0 (14):\penalty0 2934--2936,
  2003.

\bibitem[Johnson et~al.(2013)Johnson, Maznev, Cuffe, Eliason, Minnich, Kehoe,
  Torres, Chen, and Nelson]{johnson2013}
J.~A. Johnson, A.~A. Maznev, J.~Cuffe, J.~K. Eliason, A.~J. Minnich, T.~Kehoe,
  C.~M.~S. Torres, G.~Chen, and K.~A. Nelson.
\newblock Direct measurement of room-temperature nondiffusive thermal transport
  over micron distances in a silicon membrane.
\newblock \emph{Phys.~Rev.~Lett}, 110\penalty0 (2):\penalty0 025901, 2013.

\bibitem[Back et~al.(2014)Back, McCue, and Moroney]{back2014}
J.~M. Back, S.~W. McCue, and T.~J. Moroney.
\newblock Including nonequilibrium interface kinetics in a continuum model for
  melting nanoscaled particles.
\newblock \emph{Sci.~Rep.}, 4, 2014.

\bibitem[Font and Myers(2013)]{font2013}
F.~Font and T.~G. Myers.
\newblock Spherically symmetric nanoparticle melting with a variable phase
  change temperature.
\newblock \emph{J.~Nanopart.~Res.}, 15\penalty0 (12):\penalty0 2086, 2013.

\bibitem[Font et~al.(2015)Font, Myers, and Mitchell]{font2015}
F.~Font, T.~G. Myers, and S.~L. Mitchell.
\newblock A mathematical model for nanoparticle melting with density change.
\newblock \emph{Microfluid.~Nanofluid.}, 18\penalty0 (2):\penalty0 233--243,
  2015.

\bibitem[McCue et~al.(2008)McCue, Wu, and Hill]{mccue2008}
S.~W. McCue, B.~Wu, and J.~M. Hill.
\newblock Micro/nanoparticle melting with spherical symmetry and surface
  tension.
\newblock \emph{IMA J.~Appl.~Math.}, 74\penalty0 (3):\penalty0 439--457, 2008.

\bibitem[Myers and Font(2015)]{myers2015}
T.~G. Myers and F.~Font.
\newblock On the one-phase reduction of the {S}tefan problem with a variable
  phase change temperature.
\newblock \emph{Int.~Commun.~Heat Mass}, 61:\penalty0 37--41, 2015.

\bibitem[Ribera and Myers(2016)]{ribera2016}
H.~Ribera and T.~G. Myers.
\newblock A mathematical model for nanoparticle melting with size-dependent
  latent heat and melt temperature.
\newblock \emph{Microfluid.~Nanofluid.}, 20\penalty0 (11):\penalty0 147, 2016.

\bibitem[Florio and Myers(2016)]{florio2016}
B.~J. Florio and T.~G. Myers.
\newblock The melting and solidification of nanowires.
\newblock \emph{J.~Nanopart.~Res.}, 18\penalty0 (6):\penalty0 1--12, 2016.

\bibitem[Goswami and Nanda(2010)]{goswami2010}
G.~K. Goswami and K.~K. Nanda.
\newblock Size-dependent melting of finite-length nanowires.
\newblock \emph{The Journal of Physical Chemistry C}, 114\penalty0
  (34):\penalty0 14327--14331, 2010.

\bibitem[Myers(2016)]{myers2016}
T.~G. Myers.
\newblock Mathematical modelling of phase change at the nanoscale.
\newblock \emph{Int.~Commun.~Heat Mass}, 76:\penalty0 59--62, 2016.

\bibitem[Jou et~al.(2010)Jou, Casas-V{\'a}zquez, and Lebon]{jou2010}
D.~Jou, J.~Casas-V{\'a}zquez, and G.~Lebon.
\newblock \emph{Extended Irreversible Thermodynamics}.
\newblock Springer, fourth edition, 2010.

\bibitem[Cattaneo(1958)]{cattaneo1958}
C.~Cattaneo.
\newblock A form of heat conduction equation which eliminates the paradox of
  instantaneous propagation.
\newblock \emph{Compte Rendus}, 247\penalty0 (4):\penalty0 431--433, 1958.

\bibitem[Guyer and Krumhansl(1966{\natexlab{a}})]{guyer1966i}
R.~A. Guyer and J.~A. Krumhansl.
\newblock Solution of the linearized phonon boltzmann equation.
\newblock \emph{Phys.~Rev.}, 148\penalty0 (2):\penalty0 766,
  1966{\natexlab{a}}.

\bibitem[Guyer and Krumhansl(1966{\natexlab{b}})]{guyer1966ii}
R.~A. Guyer and J.~A. Krumhansl.
\newblock Thermal conductivity, second sound, and phonon hydrodynamic phenomena
  in nonmetallic crystals.
\newblock \emph{Phys.~Rev.}, 148\penalty0 (2):\penalty0 778,
  1966{\natexlab{b}}.

\bibitem[Alvarez et~al.(2009)Alvarez, Jou, and Sellitto]{alvarez2009}
F.~X. Alvarez, D.~Jou, and A.~Sellitto.
\newblock Phonon hydrodynamics and phonon-boundary scattering in nanosystems.
\newblock \emph{J.~Appl.~Phys.}, 105\penalty0 (1):\penalty0 014317, 2009.

\bibitem[Calvo-Schwarzw{\"a}lder et~al.(2018)Calvo-Schwarzw{\"a}lder, Hennessy,
  Torres, Myers, and Alvarez]{calvo2018}
M.~Calvo-Schwarzw{\"a}lder, M.~G. Hennessy, P.~Torres, T.~G. Myers, and F.~X.
  Alvarez.
\newblock A slip-based model for the size-dependent effective thermal
  conductivity of nanowires.
\newblock \emph{Int.~Comm.~Heat Mass}, 91:\penalty0 57--63, 2018.

\bibitem[Colli and Grasselli(1993)]{colli1993}
P.~Colli and M.~Grasselli.
\newblock Hyperbolic phase change problems in heat conduction with memory.
\newblock \emph{Proc.~Roy.~Soc.~Edinb.~A}, 123\penalty0 (3):\penalty0 571--592,
  1993.

\bibitem[Friedman and Hu(1989)]{friedman1989}
A.~Friedman and B.~Hu.
\newblock The {S}tefan problem for a hyperbolic heat equation.
\newblock \emph{J.~Math.~Anal.~Appl.}, 138\penalty0 (1):\penalty0 249--279,
  1989.

\bibitem[Showalter and Walkington(1987)]{showalter1987}
R.~E. Showalter and N.~J. Walkington.
\newblock A hyperbolic {S}tefan problem.
\newblock \emph{Q.~Appl.~Math.}, 45\penalty0 (4):\penalty0 769--781, 1987.

\bibitem[Glass et~al.(1991)Glass, Necati~Ozisik, McRae, and Kim]{glass1991}
D.~E. Glass, M.~Necati~Ozisik, S.~S. McRae, and W.~S. Kim.
\newblock Formulation and solution of hyperbolic {S}tefan problem.
\newblock \emph{J.~Appl.~Phys.}, 70\penalty0 (3):\penalty0 1190--1197, 1991.

\bibitem[Greenberg(1987)]{greenberg1987}
J.~M. Greenberg.
\newblock A hyperbolic heat transfer problem with phase changes.
\newblock \emph{IMA J.~Appl.~Math.}, 38\penalty0 (1):\penalty0 1--21, 1987.

\bibitem[Solomon et~al.(1985)Solomon, Alexiades, Wilson, and
  Drake]{solomon1985}
A.~D. Solomon, V.~Alexiades, D.~G. Wilson, and J.~Drake.
\newblock On the formulation of hyperbolic {S}tefan problems.
\newblock \emph{Q.~Appl.~Math.}, 43\penalty0 (3):\penalty0 295--304, 1985.

\bibitem[Liu et~al.(2009)Liu, Bussmann, and Mostaghimi]{liu2009}
H.~Liu, M.~Bussmann, and J.~Mostaghimi.
\newblock A comparison of hyperbolic and parabolic models of phase change of a
  pure metal.
\newblock \emph{Int.~J.~Heat Mass Tran.}, 52\penalty0 (5):\penalty0 1177--1184,
  2009.

\bibitem[Sadd and Didlake(1977)]{sadd1977}
M.~H. Sadd and J.~E. Didlake.
\newblock Non-{F}ourier melting of a semi-infinite solid.
\newblock \emph{J.~Heat Transf.}, 99\penalty0 (1):\penalty0 25--28, 1977.

\bibitem[Mullis(1997)]{mullis1997}
A.~M. Mullis.
\newblock Rapid solidification within the framework of a hyperbolic conduction
  model.
\newblock \emph{Int.~J.~Heat Mass Tran.}, 40\penalty0 (17):\penalty0
  4085--4094, 1997.

\bibitem[Wang and Prasad(2000)]{wang2000}
G.-X. Wang and V.~Prasad.
\newblock Microscale heat and mass transfer and non-equilibrium phase change in
  rapid solidification.
\newblock \emph{Mat.~Sci.~Eng. A-Struc}, 292\penalty0 (2):\penalty0 142--148,
  2000.

\bibitem[Deng and Liu(2003)]{deng2003}
Z.-S. Deng and J.~Liu.
\newblock Non-{F}ourier heat conduction effect on prediction of temperature
  transients and thermal stress in skin cryopreservation.
\newblock \emph{J.~Therm.~Stresses}, 26\penalty0 (8):\penalty0 779--798, 2003.

\bibitem[Ahmadikia and Moradi(2012)]{ahmadikia2012}
H.~Ahmadikia and A.~Moradi.
\newblock Non-{F}ourier phase change heat transfer in biological tissues during
  solidification.
\newblock \emph{Heat Mass Transfer}, 48\penalty0 (9):\penalty0 1559--1568,
  2012.

\bibitem[Kumar et~al.(2017)Kumar, Kumar, Katiyar, and Telles]{kumar2017}
A.~Kumar, S.~Kumar, V.~K. Katiyar, and S.~Telles.
\newblock Phase change heat transfer during cryosurgery of lung cancer using
  hyperbolic heat conduction model.
\newblock \emph{Comput.~Biol.~Med.}, 84:\penalty0 20--29, 2017.

\bibitem[Sobolev(1995)]{sobolev1995}
S.~L. Sobolev.
\newblock Two-temperature stefan problem.
\newblock \emph{Phys.~Lett.~A}, 197\penalty0 (3):\penalty0 243--246, 1995.

\bibitem[Sobolev(1997)]{sobolev1997}
S.~L. Sobolev.
\newblock Local non-equilibrium transport models.
\newblock \emph{Phys.~Usp.}, 40\penalty0 (10):\penalty0 1043--1053, 1997.

\bibitem[Sobolev(1991)]{sobolev1991}
S.~L. Sobolev.
\newblock Transport processes and traveling waves in systems with local
  nonequilibrium.
\newblock \emph{Phys.~Usp.}, 34\penalty0 (3):\penalty0 217, 1991.

\bibitem[Sobolev(1996)]{sobolev1996}
S.~L. Sobolev.
\newblock The local-nonequilibrium temperature field around the melting and
  crystallization front induced by picosecond pulsed laser irradiation.
\newblock \emph{Int.~J.~Thermophys.}, 17\penalty0 (5):\penalty0 1089--1097,
  1996.

\bibitem[Kov{\'a}cs and V{\'a}n(2015)]{kovacs2015}
R.~Kov{\'a}cs and P.~V{\'a}n.
\newblock Generalized heat conduction in heat pulse experiments.
\newblock \emph{Int.~J.~Heat Mass Tran.}, 83:\penalty0 613--620, 2015.

\bibitem[Moosaie(2008)]{moosaie2008}
A.~Moosaie.
\newblock Non-{F}ourier heat conduction in a finite medium with insulated
  boundaries and arbitrary initial conditions.
\newblock \emph{Int.~Commun.~Heat Mass}, 35\penalty0 (1):\penalty0 103--111,
  2008.

\bibitem[V{\'a}n et~al.(2017)V{\'a}n, Berezovski, F{\"u}l{\"o}p, Gr{\'o}f,
  Kov{\'a}cs, Lovas, and Verh{\'a}s]{van2017}
P.~V{\'a}n, A.~Berezovski, T.~F{\"u}l{\"o}p, Gy. Gr{\'o}f, R.~Kov{\'a}cs,
  {\'A}.~Lovas, and J.~Verh{\'a}s.
\newblock Guyer-{K}rumhansl-type heat conduction at room temperature.
\newblock \emph{arXiv preprint arXiv:1704.00341}, 2017.

\bibitem[Galenko and Danilov(1999)]{galenko1999}
P.~K. Galenko and D.~A. Danilov.
\newblock Model for free dendritic alloy growth under interfacial and bulk
  phase nonequilibrium conditions.
\newblock \emph{J.~Cryst.~Growth}, 197\penalty0 (4):\penalty0 992--1002, 1999.

\bibitem[Bender and Orszag(2013)]{bender2013}
C.~M. Bender and S.~A. Orszag.
\newblock \emph{Advanced Mathematical Methods for Scientists and Engineers I:
  Asymptotic Methods and Perturbation Theory}, chapter~9.
\newblock Springer, 2013.

\bibitem[Hinch(1991)]{hinch1991}
E.~J. Hinch.
\newblock \emph{Perturbation {M}ethods}, chapter~5.
\newblock Cambridge University Press, 1991.

\bibitem[Dong et~al.(2014)Dong, Cao, and Guo]{dong2014}
Y.~Dong, B.-Y. Cao, and Z.-Y. Guo.
\newblock Size dependent thermal conductivity of {S}i nanosystems based on
  phonon gas dynamics.
\newblock \emph{Physica E}, 56:\penalty0 256--262, 2014.

\bibitem[Ma(2012)]{ma2012}
Y.~Ma.
\newblock Size-dependent thermal conductivity in nanosystems based on
  non-{F}ourier heat transfer.
\newblock \emph{Appl.~Phys.~Lett.}, 101\penalty0 (21):\penalty0 211905, 2012.

\bibitem[Sobolev(2014)]{sobolev2014}
S.~L. Sobolev.
\newblock Nonlocal diffusion models: {A}pplication to rapid solidification of
  binary mixtures.
\newblock \emph{Int.~J.~Heat Mass Tran.}, 71:\penalty0 295--302, 2014.

\bibitem[Bolmatov et~al.(2012)Bolmatov, Brazhkin, and Trachenko]{bolmatov2012}
D.~Bolmatov, V.~V. Brazhkin, and K.~Trachenko.
\newblock The phonon theory of liquid thermodynamics.
\newblock \emph{Sci.~Rep.}, 2:\penalty0 421, 2012.

\end{thebibliography}

\end{document}